\documentclass[preprint]{aastex}

\slugcomment{{\it Accepted by the Astronomical Journal}}
\begin{document}

\title{A New Non-Parametric Approach to Galaxy Morphological Classification}

\author{Jennifer M. Lotz \altaffilmark{1}, Joel Primack \altaffilmark{1}, and
Piero Madau \altaffilmark{2}}

\altaffiltext{1}{Santa Cruz Institute of Particle Physics, 
University of California, Santa Cruz, CA 95064; jlotz@scipp.ucsc.edu, 
joel@scipp.ucsc.edu}
\altaffiltext{2}{Department of Astronomy and Astrophysics, 
University of California, Santa Cruz, CA 95064; pmadau@ucolick.org}

\begin{abstract}
We present two new non-parametric methods for quantifying galaxy morphology:
the relative distribution of the galaxy pixel flux values (the Gini coefficient or $G$) and
the second-order moment of the brightest 20\% of the galaxy's flux ($M_{20}$).
We test the robustness of $G$ and $M_{20}$ to decreasing signal-to-noise and 
spatial resolution, and
find that both measures are reliable to within 10\% for images with 
average signal-to-noise per pixel
greater than 2 and resolutions better than 1000 pc and 500 pc, respectively.
We have measured $G$ and $M_{20}$, as well as concentration ($C$), asymmetry ($A$), and
clumpiness ($S$) in the rest-frame near-ultraviolet/optical wavelengths 
for 148 bright local ``normal" Hubble type galaxies (E-Sd) galaxies,
22 dwarf irregulars, and 73 $0.05 < z < 0.25$ ultra-luminous infrared galaxies (ULIRGs).
We find that most local galaxies follow a tight
sequence in $G-M_{20}-C$, where early-types have high $G$ and $C$ and low $M_{20}$ 
and late-type spirals have lower $G$ and $C$ and higher $M_{20}$. 
The majority of ULIRGs lie above the normal galaxy $G-M_{20}$ sequence, 
due to their high $G$ and $M_{20}$ values.
Their high Gini coefficients arise from very bright nuclei, while the high second-order
moments are produced by multiple nuclei and bright tidal tails. All of these features
are signatures of recent and on-going mergers and interactions.
We also find that in combination with $A$ and $S$, $G$ is more effective than $C$ at
distinguishing ULIRGs from the ``normal" Hubble-types. 
Finally, we measure the morphologies of 49 $1.7 < z < 3.8$ galaxies from 
HST NICMOS observations of the Hubble Deep Field North.  We find that many of the $z \sim 2$
galaxies possess $G$ and $A$ higher than expected from degraded images of 
local elliptical and spiral galaxies, and have morphologies more like low-redshift
ULIRGs.
\end{abstract}

\section{INTRODUCTION}
The evolution of the physical structure of galaxies is one of the
keys to understanding  how matter in the universe assembled into the
structures we see today.  The most accessible tracer of a galaxy's
physical structure is its morphology, i.e. the organization of its 
light (stars and dust), as projected into our line of sight 
and observed at a particular wavelength. 
As we examine more distant galaxies, we find that galaxy morphologies become
increasingly chaotic. The disk and spheroidal structures abundant
in the local universe disappear at early times in the universe (e.g. Abraham
et al. 1996, Abraham \& van den Bergh 2002). The emergence of the local Hubble 
sequence of spiral and elliptical galaxies at late times is one of the predictions of the
hierarchical picture of galaxy assembly.

While the first morphological studies sought to describe the variety of galaxy
shapes and forms, the goal of present-day morphological studies 
is to tie the spatial distribution of stars to the formation history of the galaxy.  
A major obstacle to this goal has been the difficulty in quantifying 
morphology with a few simple, reliable measurements.   
One tack is to describe a galaxy parametrically, by modeling 
the distribution of light as projected into the plane of the sky
with a prescribed analytic function. 
For example, bulge-to-disk light ratios may be computed by fitting the galaxy 
with a two-component profile, where the fluxes, sizes,
concentrations, and orientations of the bulge and disk components 
are free parameters (Peng et al. 2002, Simard et al. 2002). This B/D ratio correlates with qualitative 
Hubble type classifications, although with significant scatter.  Unfortunately, there is often
a fair amount of degeneracy in the best-fitting models and B/D ratios, and 
structures such as compact nuclei, bars, and spiral arms introduce additional difficulty 
in fitting the bulge and disk components (e.g. Balcells et al 2003).
A related approach is to fit a single Sersic profile to the entire galaxy (Blanton et al. 2003a).  
Profiles with high Sersic indices are interpreted as bulge-dominated systems, while
low Sersic indices indicate disk-dominated systems.  However, not all bulges have
high Sersic index values - some are exponential in nature (Carollo 1999), so not
all objects with bulges will produce intermediate or high Sersic indices.
Both the one-component and multiple-component fitting methods
assume that the galaxy is well described by a smooth, symmetric profile 
- an assumption that breaks down for irregular, tidally disturbed, and merging galaxies.

Non-parametric measures of galaxy morphology do not assume a particular 
analytic function for the galaxy's light distribution, and therefore may be
applied to irregulars as well as standard Hubble type galaxies.  Abraham
et al. (1994, 1996) introduced the 
concentration index $C$ (which roughly correlates with a galaxy's B/D ratio)  
and Schade et al. (1995) put forward rotational asymmetry $A$ as a way to automatically 
distinguish early Hubble types (E/S0/Sa) from later Hubble types (Sb/Sc) and
classify irregular and merging galaxies. Subsequent authors modified the original definitions
to make $C$ and $A$ more robust to surface-brightness selection and centering
errors (Wu 1999, Bershady et al. 2000, Conselice et al. 2000).   The third quantity in
the ``$CAS$'' morphological classification system is a measure of a galaxy's residual clumpiness 
$S$, which is correlated with a galaxy's color and star-formation rate 
(Isserstedt \& Schindler 1986; Takamiya 1999; Conselice 2003).  
Other more computer-intensive approaches to galaxy classification such as artificial neural networks
and shapelet decomposition have also been applied to
local and distant galaxies.  Artificial neural networks are trained by an astronomer on a set of 
galaxies of known morphological type and use a combination of size, surface-brightness, concentration, and color 
to classify galaxy types (Odewahn et al. 1996, Naim et al. 1997).   ``Shaplets''  deconstruct
each galaxy's image into a series of Hermite polynomials (Refregier 2003,  Kelly \& McKay 2004).
The eigen-shapes produced by shapelet decomposition are often difficult to interpret by themselves,
and the additional step of principle component analysis is performed to classify galaxies. 

While $CAS$ is perhaps the most straightforward of the non-parametric 
methods, it is not without its weaknesses.  Because concentration is measured within several
circular apertures about a pre-defined center, it implicitly assumes circular symmetry, 
making it a poor descriptor for irregular galaxies.  Asymmetry is more sensitive to 
merger signatures than concentration, but not all merger remnant candidates are highly asymmetric, 
and not all asymmetric galaxies  are mergers (e.g. dusty edge-on spirals).  Finally, the clumpiness determination 
requires one to define a galaxy smoothing length, 
which must be chosen carefully to avoid systematic effects dependent on a galaxy image's point spread function (PSF),
pixel scale, distance, and angular size.  Also, the bulges of highly concentrated
galaxies give strong residuals which are not due to star-forming regions and must be masked
out when computing $S$.  

In this paper, we examine two new non-parametric ways of quantifying galaxy morphology which 
circumvent some of the problems with the ``$CAS$'' system.  We use the Gini coefficient,
a statistic used in economics to describe the distribution of wealth within a
society. It was first adapted for galaxy morphology classification
by Abraham et al. (2003) to quantify the relative distribution of flux within the 
pixels associated with a galaxy.  It is correlated with concentration, but 
does not assume that the brightest pixels are in the geometric center of the galaxy image.
We also define a new indicator, $M_{20}$, which describes the second-order moment of the brightest
20\% of the galaxy.  While similar to the concentration index, $M_{20}$ is
more sensitive to merger signatures like multiple nuclei and does not impose circular 
symmetry.  In \S 2, we modify Abraham's definition of the Gini coefficient in order to make
it applicable to distant galaxies and we define $M_{20}$. 
In \S 3, we test the robustness of these statistics to decreasing $S/N$ and  
resolution, and find that at average $S/N$ per galaxy pixel $> 2$ and spatial resolutions $<$ 500 pc, 
they are reliable to within 10\%.  We also compare the robustness of $G$ and $M_{20}$ to $CAS$.
In \S 4, we compare the ability of $G$ and $M_{20}$ to classify
local Hubble-type and merging galaxies to the $CAS$ system. 
Finally in \S 5, we examine the near-ultraviolet/optical morphologies of 
49 $ 1.7 < z < 3.8$ Lyman break galaxies
and attempt to classify these LBGs as ellipticals, spirals, or merger candidates.

\section{MEASURING GALAXY MORPHOLOGIES}

\subsection{The Gini Coefficient}
The Gini coefficient is a statistic based on the Lorenz curve, the 
rank-ordered cumulative distribution function of a population's wealth or, in this
case, a galaxy's pixel values (Abraham et al. 2003).  
The Lorenz curve is defined as 
\begin{equation}
L(p) = \frac{1}{\bar{X}} \int_0^p{F^{-1}(u) du}
\end{equation}
where $p$ is the percentage of the poorest citizens or faintest pixels, 
F(x) is the cumulative distribution function, and $\bar{X}$ is the
mean over all (pixel flux) values $X_i$ (Lorenz 1905).   
The Gini coefficient is the ratio of the area between 
the Lorenz curve and the curve of ``uniform equality'' where $L(p) = p$ 
(shaded region, Figure 1) to the area under the curve of uniform equality ($= 1/2$). 
For a discrete population, the Gini coefficient 
is defined as the mean of the absolute difference between all $X_i$:
\begin{equation}
G = \frac{1}{2 \bar{X} n (n-1)} \sum^n_{i=1} \sum^n_{j=1} | X_i - X_j |
\end{equation}
where $n$ is the number of people in a population or pixels in a galaxy. 
In a completely egalitarian society, $G$ is zero, and if one individual has all the wealth, $G$ is unity.  
A more efficient way to compute $G$ is to first sort $X_i$ into increasing order
and calculate
\begin{equation}
G = \frac{1}{\bar{X} n (n-1)} \sum^n_i (2i - n -1) X_i
\end{equation}
(Glasser 1962).

For the majority of local galaxies, the Gini coefficient is correlated with the concentration
index, and increases with the fraction of light in a compact (central) component.
In a study of 930 SDSS Early Data Release galaxies, Abraham et al. (2003)
found $G$ to be strongly correlated with both concentration and surface brightness.
However, unlike $C$, $G$ is independent of the large-scale spatial distribution of 
the galaxy's light.  The correlation between $C$ and $G$ exists because highly concentrated
galaxies have much of their light in a small number of pixels.  High $G$ values may
also arise when very bright galaxy pixels are not found in the center of a bulge.
Therefore $G$ differs from $C$ in that 
it can distinguish between galaxies with shallow light profiles
(which have both low $C$ and $G$)  and galaxies where much of the flux is located in a 
few pixels not at the projected center (which have low $C$ but high $G$). 

In practice, the application of the Gini coefficient to galaxy observations 
requires some care. One must have a consistent definition of the pixels belonging to the galaxy 
in order to measure the distribution of flux within those pixels and compare that distribution
to other galaxies. The inclusion of ``sky'' pixels will systematically increase $G$, while
the exclusion of low-surface brightness ``galaxy'' pixels will 
systematically decrease $G$.   Abraham et al. (2003) measure $G$ for galaxy pixels which lie above
a constant surface-brightness threshold.  This definition makes the direct comparison between
high-redshift galaxies and the local galaxy population difficult because of the $(1+z)^4$ 
surface-brightness dimming of distant galaxies.  Therefore, we attempt to create a 
segmentation map of the galaxy pixels in a way that is insensitive to surface-brightness dimming. 
The mean surface brightness $\mu(r_p)$ at the Petrosian radius $r_p$ is used to set the flux threshold 
above which pixels are assigned to the galaxy. The Petrosian radius is the radius $r_p$ at which 
the ratio of the surface brightness at $r_p$ to the mean surface brightness within $r_p$ is equal
to a fixed value, i.e.
\begin{equation}
\eta = \frac{\mu(r_p)}{\bar{\mu}(r<r_p)}
\end{equation}
where $\eta$ is typically set to 0.2 (Petrosian 1976).  
Because the Petrosian radius is based on a curve of growth, it is largely insensitive 
to variations in the limiting surface brightness and  $S/N$ of the observations. This revised definition 
should allow better comparison of $G$ values for galaxies with varying surface brightnesses, distances, 
and observed signal-to-noise.. 

The galaxy image is sky-subtracted and any background galaxies, foreground stars, or cosmic rays
are removed from the image. The mean ellipticity and position angle of the galaxy is measured using
IRAF task ellipse.  The Petrosian ``radius'' (or semi-major axis length) 
is measured for increasing elliptical apertures, rather than circular apertures.  While
the Petrosian radius determined by the curve of growth within circular apertures is similar to  
that determined from elliptical apertures for most galaxies, 
elliptical apertures more closely follow the galaxy's true 
light profile and can produce very different $r_p$ values for edge-on spirals. 
To create the segmentation map,  the cleaned
galaxy image is first convolved with a Gaussian with $\sigma = r_p/5$. 
This step raises the signal of the galaxy pixels above the background noise,
making low-surface brightness galaxy pixels more detectable.
Then the surface brightness ${\mu}$ at $r_p$ is measured and pixels in the smoothed image 
with flux values $\geq  {\mu}(r_p)$ and less than 
10 $\sigma$ from their neighboring pixels are assigned to the galaxy.  
The last step assures that any remaining cosmic rays or spurious noise pixels in the 
image are not included in the segmentation map.
This map is then applied to the cleaned but unsmoothed image, and the pixels assigned
to the galaxy are used to compute the Gini coefficient.  

Even when the pixels assigned to a galaxy are robustly determined, 
the distribution of flux within the pixels will depend on the signal-to-noise
ratio ($S/N$) as noise smears out the flux distribution in the faintest pixels.  
This is illustrated in the left of Figure 2 by adding increasing Poisson sky noise to the 
S0 galaxy NGC4526 image, and recalculating the segmentation map and Gini coefficient.
We define the average signal-to-noise per galaxy pixel  $<S/N>$ as
\begin{equation}
<S/N> = \frac{1}{n} \sum^i_n \frac{ S_{i}}{\sqrt{ \sigma_{sky}^2 + S_{i}}}
\end{equation}
where $S_i$ is  pixel $i$'s flux, $\sigma_{sky}$ is the sky noise, and $n$ is the
number of galaxy pixels in the segmentation map.
As $<S/N>$ decreases, 
the distribution of measured flux values in the faintest pixels becomes broader.
The measured Gini coefficient increases because low surface-brightness galaxy
pixels are scattered to flux values below the mean sky level, resulting in
negative flux levels for the faintest pixels assigned to the galaxy by our smoothed 
segmentation map.    
We note that, while the Poisson noise redistributes all the pixel flux values, 
the effects are significant only for pixels with intrinsic flux values $\leq 3 \sigma_{sky}$.
Therefore, as a first order correction, we compute the Gini coefficient of
the distribution of {\it absolute} flux values:
\begin{equation}
G = \frac{1}{\bar{|X|} n (n-1)} \sum^n_i (2i - n -1) |X_i|
\end{equation}
Low-surface brightness 
galaxy pixels with flux values scattered below the sky level are reassigned
positive values (right of Figure 2). This correction recovers the ``true'' Gini coefficient to within 10\% for
images with $S/N > 2$; at very low $S/N$ values, even the brightest galaxy pixels
are strongly affected by noise and the Gini coefficient is not recoverable.
In Figures 3-4, we show the final segmentation maps used to compute the Gini 
coefficient as contour maps for eight galaxies of varying morphological type (Table 1).

\subsection{The Moment of Light}
The total second-order moment $M_{tot}$ 
is the flux in each pixel $f_i$ multiplied by the squared distance to the center of the galaxy, 
summed over all the galaxy pixels assigned by the segmentation map:
\begin{equation} 
M_{tot} = \sum_i^n M_i = \sum_i^n f_i \cdot ((x_i - x_c)^2 + (y_i - y_c)^2)
\end{equation}
where $x_c, y_c$ is the galaxy's center. The center is computed by finding $x_c, y_c$ 
such that $M_{tot}$ is minimized.

The second-order moment of the brightest regions of the galaxy  
traces the spatial distribution of any bright nuclei, bars, 
spiral arms, and off-center star-clusters. 
We define $M_{20}$ as the normalized second order moment of the brightest 20\% of
the galaxy's flux.  To compute $M_{20}$, we rank-order the galaxy pixels by flux, 
sum $M_i$ over the brightest pixels until the sum of the brightest pixels equals 20\% of the 
total galaxy flux, and then normalize by $M_{tot}$:
\begin{eqnarray}
M_{20} \equiv {\rm log10}\left(\frac{\sum_i M_i}{M_{tot}}\right) & {\rm while } & \sum_i f_i <  0.2 f_{tot}
\end{eqnarray}
Here $f_{tot}$ is the total flux of the galaxy pixels identified by the
segmentation map and $f_i$ are the fluxes for each pixel $i$, order such that $f_1$ is the brightest pixel, 
$f_2$ is the second brightest pixels, and so on. The normalization by $M_{tot}$ 
removes the dependence on total galaxy flux or size. We find that defining $M$ with brighter flux
thresholds (e.g. 5\% of f$_{tot}$) produce moment values that are unreliable at 
low spatial resolutions (\S 2.3), while lower flux threshold lead to a less discriminating statistic.

While our definition of $M_{20}$ is similar to that of $C$, it differs in two important
respects.  Firstly,  $M_{20}$ depends on $r^2$, and is more heavily weighted by the 
spatial distribution of luminous regions.  Secondly, unlike $C$, $M_{20}$ is not
measured within circular or elliptical apertures, and the center of the galaxy is a free
parameter.  We shall see in \S 3 that these differences make $M_{20}$ more sensitive than $C$ to merger signatures 
such as multiple nuclei.  In Figures 3 and 4, we display the segmentation maps and the
regions containing the brightest 20\% of the flux for the eight test galaxies.

\subsection{ Concentration, Asymmetry, and Smoothness}
Concentration is defined in slightly different ways by different authors, but the basic
function measures the ratio of light within a circular or elliptical inner aperture 
to the light within an outer aperture.  We adopt the Bershady et al. (2000) definition as the ratio of the 
circular radii containing 20\% and 80\% of the ``total flux'' :
\begin{equation}
C = 5\ \rm{ log10}\left(\frac{r_{80}}{r_{20}}\right)
\end{equation}
where $r_{80}$ and $r_{20}$ are the circular apertures containing 80\% and 20\%
of the total flux, respectively.   For comparison to the most recent
studies of galaxy concentration, we use Conselice's (2003) definition of the total flux as
the flux contained within 1.5 $r_p$ of the galaxy's center (as opposed to 
Bershady's definition as the flux contained within 2 $r_p$).  For 
the concentration measurement, the galaxy's center is that determined by
the asymmetry minimization (see below).  In Figures 3-4, we over-plot $r_{80}$ and $r_{20}$ 
for eight galaxies of varying morphological type 
in the far left-hand panels. 

The asymmetry parameter $A$ quantifies the degree to which the light of a galaxy is 
rotationally symmetric.  
$A$ is measured by subtracting the galaxy image rotated by 180 degrees from the
original image (Abraham et al. 1995, Wu 1999, Conselice et al. 2000). 
\begin{equation}
A = \sum_{i,j} \frac{ | I(i,j) - I_{180}(i,j)|}{|I(i,j)|} - B_{180}
\end{equation}
where $I$ is the galaxy's image and $I_{180}$ is the image rotated by 180 about
the galaxy's central pixel, and $B_{180}$ is the average asymmetry of the background. 
$A$ is summed over all pixels within 1.5 $r_{p}$ of the galaxy's center.  The
central pixel is determined by minimizing $A$.  The asymmetry due to the noise must be 
corrected for, and it is impossible to reliably 
measure the asymmetry for low $S/N$ images. In Figures 3-4, we display the
residual $I - I_{180}$ image and the 1.5 $r_p$ aperture 
in the second column.  Objects with very smooth elliptical light profiles have a high
degree of rotational symmetry.
Galaxies with spiral arms are less symmetric, while extremely irregular and 
merging galaxies are often (but not always) highly asymmetric.     

The smoothness parameter $S$ has been recently developed by Conselice (2003), inspired by
the work of Takamiya (1999), in order to quantify the degree of small-scale structure.  
The galaxy image is smoothed by a boxcar of given width and then
subtracted from the original image.   The residual is a measure of the clumpiness due to
features such as compact star clusters.  
In practice, the smoothing scalelength is chosen to be a fraction of the Petrosian radius.
\begin{equation}
S =  \sum_{i,j}\frac{ | I(i,j) - I_S(i,j)| } {|I(i,j)|} - B_S 
\end{equation}
where $I_S$ is the galaxy's image smoothed by a boxcar of width 0.25 $r_p$, and
$B_S$ is the average smoothness of the background.  Like $A$, $S$ is summed over the
pixels within 1.5 $r_p$ of the galaxy's center.  However, because the central regions
of most galaxies are highly concentrated, the pixels within a circular aperture
equal to the smoothing length 0.25 $r_p$ are excluded from the sum.
In Figures 3-4, we display the residual $I- I_S$ images, and the 0.25 and 1.5 $r_p$ apertures
in the third column.  $S$ is correlated with recent star-formation
(Takamiya 1999, Conselice 2003).  However, because of its strong dependence on resolution,  
it is not applicable to  poorly resolved and distant galaxies.

\section{RESOLUTION AND NOISE EFFECTS}
In order to make a fair comparison of the measured morphologies of different galaxies, 
we must understand how noise and resolution affect $G$ and $M_{20}$.  This is particularly important
when comparing local galaxies to high-redshift galaxies, as the observations of
distant galaxies are generally of lower signal-to-noise and resolution than those
of local galaxies.  We have defined $G$ and $M_{20}$  in the previous 
sections in an attempt to minimize systematic offsets with noise and resolution.
Nevertheless, any measurement is ultimately limited by the $S/N$ of the observations.
Also, the PSF and finite pixel size of the images may introduce increasing uncertainties 
to the morphologies as the resolution decreases and small-scale structures are washed out. 

We have chosen eight galaxies of varying morphological type (Figures 3 and 4; Table 1)
to independently test the effects of decreasing $S/N$ per pixel and physical resolution (pc per pixel) 
on the measurements of $G$, $M_{20}$, $C$, $A$, and $S$.  
For the $S/N$ tests, random Poisson noise maps of increasing variance were added to 
the original sky-subtracted image. For each noise-added image, we measured $r_{p}$, 
created a new segmentation map,  
measured $<S/N>$ for galaxy pixels assigned by the segmentation map, 
and measured $G$, $M_{20}$, $C$, $A$, and $S$.  Noisy galaxy images were created and measured 20 times at each $S/N$ level,
and the mean changes in the morphological values with $<S/N>$ are plotted in Figure 5. 
To simulate the effect of decreasing resolution, we re-binned the galaxy images to increasingly
large pixel sizes. Re-binning the original galaxy images increases the $S/N$ per pixel, 
so additional Poisson sky noise ($\sigma_{sky}$) was added to the re-binned image such that average
$<S/N>$ was kept constant
with decreasing resolution.  Again, we measured $r_{p}$, created a segmentation map, and computed
the average change $G$, $M_{20}$, $C$, $A$, and $S$ with resolution for 20 simulations at each resolution step
(Figure 6).

We find that $G$, $M_{20}$, and $C$ are reliable to within $\sim$ 
10\% ($\Delta \leq$ 0.05, 0.2, and 0.3 respectively)
 for galaxy images with $<S/N> \geq 2$.   
$A$ systematically decreases with $<S/N>$, but generally shows offsets less than 0.1 at $<S/N> \geq 5$.
$S$ also systematically decreases with $<S/N>$, and has decrements less than 0.2 at $<S/N> \geq 5$.
Decreasing resolution, however, has much stronger effects on the morphology measurements.  
$C$ and $M_{20}$ show systematic offsets greater than $\sim$ 15\% 
($\Delta \geq 0.5$ and 0.3, respectively) at resolution scales
worse than 500 pc, as the cores of the observed galaxies become unresolved.  
$G$, $A$, and $S$, on the other hand, are relatively stable to decreasing spatial resolution 
down to 1000 pc.   As a galaxy's image becomes less resolved, the observed curve of growth changes 
resulting in larger $r_p$ values, and therefore producing slightly higher $G$ values as 
the segmentation map grows accordingly. At the lowest resolutions, the observed biases
in $C$, $A$ and $S$ appear to be a function of Hubble type:  the E-Sbc galaxies are biased to higher $A$
and $S$ and lower $C$, while both the Sd and mergers are biased toward lower $A$ and 
the merger remnants are biased to higher $C$.  On the other hand, on the Sc and Sd galaxies show
$G$ offsets $>$ 20\% ($\Delta \sim 0.1$) at resolutions between 1000 and 2000 pc.

\section{LOCAL GALAXY MORPHOLOGIES}

\subsection{Frei and SDSS Local Galaxy Samples}
We have measured $G$, $M_{20}$, $C$, $A$, and $S$ at both $\sim 4500$\AA\ and $\sim 6500$\AA\
for 104 local galaxies taken from the Frei et al. (1996) catalog.  
The Frei catalog galaxies are a representative sample of bright, well-resolved, 
Hubble-type galaxies (E-S0-Sa-Sb-Sc-Sd), and have been used as morphological standards by 
a number of authors (Takamiya 1999; Wu 1999; Bershady et al. 2000; Conselice et al. 2000; 
Simard et al. 2002). The galaxies were observed by Frei et al. (1996) with either the 1.5 meter telescope at 
Palomar Observatory or the 1.1 meter telescope at Lowell Observatory. 
 The Palomar images were taken in the Thuan-Gunn $g$ and $r$ filters
($\lambda_{eff} =$ 5000\AA , 6500\AA ) 
at plate scale = 1.19~\arcsec\ per pixel and typical PSF FWHM $\sim$ 2-3~\arcsec.  The Lowell 
images were taken in the $B_J$ and $R$ pass-bands ($\lambda_{eff}$ = 4500\AA , 6500\AA)  
at a plate scale = 1.35~\arcsec\ per pixel and typical PSF FWHM $\sim$~3-5~\arcsec.  
In Table 2, we give $G$, $M_{20}$, $C$, $A$, and $S$ as measured in
$R$/$r$ and $B_J$/$g$ for each of the galaxies. 

We have also obtained the images of 9 Frei galaxies and 44 other galaxies selected by
their $u$-band brightness ($u < 14$) 
from the SDSS Data Release 1 database (Abazajian et al. 2003).  The morphologies of the SDSS sample 
were measured in the $u$, $g$ and $r$-bands ($\lambda_{eff}$ = 3600\AA, 4400\AA, and 6500\AA\
respectively; Table 3). The SDSS plate scale is 0.4\arcsec\ per pixel and 
the $r$-band PSF FWHM values are typically  $\sim$~1.3-1.8 \arcsec (Stoughton et al. 2002).
We find that the mean absolute difference between the SDSS and Frei observations are :
\begin{eqnarray}
\begin{array}{llllll}
\delta G = 0.02 & \delta M_{20} = 0.12 & \delta C =  0.11 & \delta A = 0.04 & \delta S = 0.09, & r/R\ {\rm band}\\
\delta G = 0.02 & \delta M_{20} = 0.11 & \delta C =  0.14 & \delta A = 0.05 & \delta S = 0.14, & g/B\ {\rm band} \\
\end{array}
\end{eqnarray}
In addition, we have analyzed $B$-band images of 22 nearby dwarf irregular galaxies from the Van Zee (2001)
sample (Table 4).  We have selected galaxies from the original Van Zee sample with 
minimal foreground star contamination and $<S/N>$ $\geq$ 2.  These images were obtained at the
Kitt Peak 0.9 m telescope and have PSF FWHM $\sim$~1.4-2.3~\arcsec and a plate scale=0.688~\arcsec. 

In Figures 7-8, we examine the dependence of $C$, $A$, $S$, $G$, and $M_{20}$ on the observed 
near-ultraviolet/optical wavelength.  
For the majority of galaxies, the difference between the observed morphologies
at $\sim$ 4500\AA\ ($B/g$)and 6500\AA\ ($R/r$) are comparable to the observational offsets between
the SDSS and Frei observations of the same galaxies in the same bandpass.  
The observed changes in $C$, $G$, and $M_{20}$ from $\sim$ 3600\AA\ ($u$) to $\sim$ 6500\AA\ ($r$)
are also consistent with observational scatter.  The SDSS $u$-band observations often have 
too low $S/N$ to obtain reliable asymmetries.  This may also produce the increased
scatter in $S$. Nevertheless, late-type galaxies generally have higher clumpiness values
and slightly higher $M_{20}$ values at 3600 \AA\ than 6500 \AA.
A handful of galaxies (many of which are edge-on spirals) show much larger morphological changes at bluer wavelengths.
The S0 galaxy UGC1597 has an obvious tidal tail, and it has higher $g$-band $A$, $S$, and $G$ values 
and a lower $g$-band $C$.  Several mid-type spirals have significantly higher $M_{20}$ values in
$B/g$ than in $R/r$.  These include NGC3675, an Sb with prominent dust features, and NGC 5850, an Sb
with a star-forming ring. 

Previous studies have noted small offsets in concentration and asymmetry 
from $U$ and $B$ to $R$, with much stronger shifts at wavelengths $\leq$ 2500\AA\ (Brinchmann et al. 1998,
Conselice et al. 2000, Kuchinski et al. 2001).   We see similar trends of slightly higher  $B/g$
asymmetries for late-type spirals ($\sim 0.05$) and lower $B/g$ concentrations for most galaxies ($\sim 0.1$).
However, given that these trends are smaller than the difference between different observations 
of the same galaxy at the same wavelength, we conclude that morphological K-corrections to $C$ and $A$ 
are not very substantial for most normal galaxies observed redward of rest-frame $\sim 3500-4000$ \AA .   
The late-type spirals show small but systematic trends of stronger clumpiness and higher second order
moments at bluer wavelengths. 

In Figure 9, we examine the $G-M_{20}$ morphologies of local galaxies observed in both
the $R/r$ and $B/g$-bands.  The distribution of local galaxies is very similar at both
wavelengths, with E/S0s showing high $G$ and low $M_{20}$ values, Sa-Sbc at intermediate 
$G$ and $M_{20}$ values, and most late-type spirals and dI with low $G$ and higher $M_{20}$ values.
Most edge-on galaxies (barred symbols) show $G$ and $M_{20}$ values consistent with the mean
values for their Hubble type.  One notable exception is the S0 NGC4710, which has
a prominent dust lane and $G=0.50$, $\sim$ 0.1 lower than for other E/S0s.
The majority of local galaxies lie below the rough dividing line plotted in Figure 9.
Four out of the 22 dIs lie above this line.  Two of these
are classified as star-bursting dwarfs (UGC11755 and UGCA439), and a third has the
bluest $U-B$ color gradient in the sample (UGC5288; Van Zee 2001).  The other outliers are UGC10991
which appears to have a tidal tail and star-forming knots, and UGC10310  
which has two very bright knots in its outer arms that may be foreground stars. 
As we discuss in the next section, most ULIRGs lie above this dividing line. 
While a few truly star-bursting dIs are $\sim 0.04$ in G above the normal galaxy
sequence at blue wavelengths, it appears that dIs will not seriously contaminate the  
merger/interacting galaxies classified by $G-M_{20}$. 

\subsection{Merger Indicators}
One of the primary goals of morphological studies is to quantitatively identify
interacting and merging galaxies.   Towards this end, Abraham (1996) and
Conselice (2000, 2003) have used combinations of concentration, asymmetry, and
smoothness to roughly classify ``normal'' galaxies as early and late-types, 
as well as to distinguish mergers from these normal types.  Abraham (2003) also found that
for a large sample of normal galaxies, the Gini coefficient is strongly correlated
with concentration, color, and surface brightness, and therefore may be as efficient as
concentration at quantifying galaxy morphologies.  Here we compare the effectiveness 
of our definition of the Gini coefficient (Equation 6)
to $C$ at classifying local galaxy types and identifying merger candidates.
We also expect that $M_{20}$ will be strongly correlated to $C$, due to their similar
definitions, and therefore examine the $G-M_{20}$ correlation and compare it to the $C-G$
relation found by Abraham et al. (2003).

In Figures 10-14, we compare the $R/r$-band morphological distributions of our local galaxy sample to 
archival HST WFPC2 F814W observations of 73 ultra-luminous infrared galaxies with
$11.5 \leq$ log($L_{FIR}/L_{\odot}$) $\leq 12.5$  and $<S/N> \geq 2 $ (ULIRGs; Borne et al. 2000, HST
Cycle 6 program 6346, Table 5).  ULIRGs often show morphological signatures of on-going or recent merger events in the form of high
asymmetries, multiple nuclei, and tidal tails (Wu et al. 1998, Borne et al. 2000, Conselice et al. 2000, Cui et al. 2001).  
We have divided the ULIRG sample into objects with ``single'', ``double'', or ``multiple'' nuclei as 
classified by Cui et al. 2000 by counting the number of surface brightness peaks with FWHM $>$ 0.14~\arcsec\ and 
$M_I < -17.0$ separated by less than 20 kpc projected.  We also identify ULIRGs in projected pairs as IRAS sources with
projected separations greater than 20 kpc and less than 120 kpc.
The ULIRG sample has a mean redshift of $\sim 0.2$, 
therefore the F814W bandpass ($\lambda_{eff} =$ 8200\AA) samples the rest-frame light at $\sim 6800$\AA.  
Given the 0.14~\arcsec\ PSF of the WF camera, ULIRGs at $z < 0.25$ are spatially resolved to better than $\sim$ 500 pc, 
and may be directly compared to the local galaxy $r/R$-band observations. 

Most ULIRGs lie above the $G-M_{20}$ correlation for normal galaxies (Figure 10, bottom panels), 
while many ULIRGs overlap with the $C-G$ and $C-M_{20}$ correlations for normal galaxies (Figure 11).  
Normal local galaxies also segregate more cleanly from the ULIRGs sample in
$G-A$ and $G-S$ than $C-A$ and $C-S$ (Figures 12-13).  In particular, the Gini coefficient of edge-on spirals 
galaxies is more consistent with the values obtained for face-on spirals. 
Also, ULIRGs with double or multiple nuclei generally have higher Gini coefficients 
relative to their concentrations than most normal galaxies.   $G-M_{20}$ is  slightly
less effective at identifying single-nuclei ULIRGs
than  $G-A$ and $G-S$; however, $M_{20}$ is a more robust indicator at 
low $S/N$ than $A$ and $S$ and at low resolution than  $S$ (Figures 5-6), and therefore may be 
applied to fainter galaxy populations.   We find that $M_{20}$ in combination with
$C$, $A$, and $S$ is not effective at separating the 
ULIRGs from the normal galaxy population (Figures 12 and 14).
In Table 6, we give the results of a series of two-dimensional Kolmogorov-Smirnov (KS) 
tests (Fasano \& Franceschini 1987)
applied to the ULIRGs and $R/r$-band normal galaxy observations for each combination
of $C$, $A$, $S$, $G$, and $M_{20}$.  For all the tests, the probability that the
ULIRGs and normal galaxies are drawn from the same parent sample is less than 
$10^{-6}$.   

While the ULIRG population as a whole occupies a different region of $C-A-S-G-M_{20}$ space than our
SDSS and Frei galaxy samples, we also find significant differences between ULIRGs in well-separated pairs, ULIRGs with
single nuclei, and ULIRGs with double or multiple nuclei (Table 6).  ULIRGs in pairs show the
smallest offsets from the normal galaxy sample. Double and multi-nuclei ULIRGs show the greatest
changes in morphology, with typically large $M_{20}$ and $A$ values.  Single-nucleus ULIRGs
appear similar to paired ULIRGs, but can also have higher $G$ and $C$.  Two dimensional KS tests show that 
the multi- and double-nuclei ULIRGs are distinct from the single-nucleus ULIRGs
and paired ULIRGs with greater than 97\% and  90\% confidence, respectively. 
The multi- and double-nuclei ULIRGs have a greater than 5\% probability of being drawn from the same sample, while
single-nucleus ULIRGs and ULIRGs in pairs have a greater than 12\% probability of being drawn from the same
sample. 

\section{GALAXY MORPHOLOGIES AT REDSHIFT $>2$}
One of the major successes of the hierarchical paradigm of galaxy formation has been
the discovery of large fractions of morphologically-irregular galaxies at $z > 1$ 
(e.g. Driver et al. 1995; Abraham et al. 1996; Odewahn et al. 1996; Abraham \& van den Bergh 2001).  
Many of these galaxies are excellent merger candidates, and
suggest merger fractions between 25-40\% at $0.5 < z < 2$.  However, morphological studies of the most distant
galaxies - the Lyman-break galaxies (LBGs) - have produced confusing and conflicting conclusions.
Initial HST WFPC2 observations of the rest-frame far-ultraviolet morphologies of 20 $z>3$ galaxies found
that they possessed one or more compact ``cores'' with sizes similar to present-day spiral bulges
(Giavalisco et al. 1996). More recent ACS observations of large numbers of $2 < z < 6$ LBGs have
confirmed ultraviolet half-light radii between 1.5 and 3.5 kpc and concentrations similar to
local bulges and ellipticals (Ferguson et al. 2003).  However, these LBGs have an ellipticity 
distribution more like disk galaxies than ellipsoids, leading to the conclusion that LBGs are drawn
from a mixture of morphological types.  Rest-frame optical observations in the near-infrared with NICMOS
have shown that the observed LBG morphologies are not a strong function of 
wavelength (Papovich et al. 2001; Dickinson 1999), and that LBGs have internal far-UV - optical color dispersions 
much smaller than $z \sim 1$ galaxies (Papovich 2002).   LBGs are significantly bluer
than local galaxies, and it is likely that their ultraviolet and optical morphologies are
dominated by young stars. Their small sizes, high concentrations, and high star-formation rates 
suggest that many are precursors to local spiral bulges. 
However, surface-brightness dimming may prevent the detection of faint tidal tails and
some appear to possess multiple nuclei.  In a recent study of the optical morphologies
of the Hubble Deep Field North galaxies, Conselice et al. (2003) found that  7 out of 18  $z<3$, $M_B < -21$
galaxies possess corrected asymmetries greater than 0.35, implying that up to 50\% are recent mergers.
However, as we found in \S 4, asymmetry is not as sensitive
by itself at detecting merger remnants as it is in combination with $C$ or $G$.  Here we re-examine the
optical morphologies of the HDFN high-redshift galaxy sample using $C$, $G$, and $M_{20}$, 
and we attempt to classify these galaxies as ellipticals, disks, or recent mergers.

The Hubble Deep Field North has 27 spectroscopically-confirmed high-redshift galaxies and 70 additional 
candidates with $1.7< z < 4$ and $H<25.0$ (Papovich et al. 2001 and references therein) .   
At these redshifts, the near-ultraviolet and optical regions of the galaxies spectral energy
distributions have been shifted to redward of 1 $\mu$m, and therefore require
infrared observations to directly compare their morphologies to the rest-frame near-UV/optical morphologies of
local galaxies. The HDFN has been observed with the NICMOS camera 3 in the F110W ($J$) and F160W ($H$) band-passes
($\lambda_{eff} = 1.1 \mu$m, $1.6\mu$m) down to a 10 $\sigma$ limiting magnitude of 26.5 (Dickinson 1999, 
HST Cycle 7 program 7817).  
Most of the HDFN LBGs are fainter than $H= 23.0$; therefore, to increase their signal-to-noise per pixel,
we have measured the morphologies of the LBG sample in a summed F110W and F160W image.
The effective central wavelength of the summed LBG observations is $\sim 1.3\mu$m. 
Galaxies at $z \sim 2$ and 3 are observed at rest-frame wavelengths 4300\AA\ and 3250\AA\, 
respectively.  Out of our initial sample of 97 $H < 25$ $1.7< z \leq 3.8$ galaxies, 
33 galaxies with $1.7 < z < 2.3$ and 16 galaxies with $2.3 \leq z \leq 3.8$ 
have $<S/N> > 2.0$ (Table 7).  We also give estimates of the
rest-frame $M_B$ in AB magnitudes, computed by interpolating between the $J$, $H$, and
$K_s$ fluxes and assuming $H_0 = 70$ km s$^{-1}$ Mpc$^{-1}$, $\Omega_\Lambda$ = 0.7, $\Omega_m$ = 0.3
cosmology.

The NICMOS images offer the highest available resolution at near-UV/optical wavelengths for these galaxies.
Nevertheless, the physical resolution of the $z > 2$ galaxies is significantly worse than that for the
local galaxy images.  The dithered NIC3 observations have a pixel scale = 0.08\arcsec\ per pixel and
a PSF FWHM = 0.22 \arcsec. At $z \geq 2$ this corresponds to a physical pixel scale of $\sim 670$ pc 
and PSF FWHM $\sim$ 1.8 kpc.  Our simulations in \S 3 showed that these resolutions produce
strong biases in the measured morphologies which are often a function of morphological type.  The
well-defined correlations of local galaxy morphologies are likely to change significantly 
with these biases.  Therefore we compare the LBG morphologies to local galaxy images
which have been measured from degraded $u$-band and $B/g$-band images.  
The galaxies are selected to lie in the same $(M_B - M^*)$ range (Figure 15), 
assuming $M^* = -20.1$ locally (Blanton et al. 2003b) and $M^* = -22.9$ at $z \geq 2$ 
(Shapley et al. 2001).  This selection tests a ``passive'' evolutionary scenario, in which
the local galaxies were brighter in the past but did not evolve morphologically.
We select local galaxies observed in $B/g$ with $3.5 \geq (M_B - M^*) \geq 1.0$ to compare
to a similarly selected $z \sim 2$ sample, and local galaxies observed in $u$ with
$2.5 \geq (M_B - M^*) \geq -0.5$ to compare to the $z \sim 3$ sample.

The local galaxies images were first deconvolved in the standard way in IDL: we
divide the Fourier transform of the image by the Fourier transform of the PSF, and compute
the inverse Fourier transform of the result.  Next they were re-binned to the pixel scale of galaxies
observed at $z=2$ (670 pc per pixel) or $z=3$ (616 pc per pixel), and convolved with the NIC3 PSF 
(FWHM = 0.22\arcsec = 2.75 pixels).  The galaxy fluxes were scaled to the count rate for an $M^*$ galaxy
at $z=2$ or $z=3$ observed by NICMOS in $F110W$ + $F160W$.  Finally, a blank region of NICMOS HDFN combined $F110W$ 
$+$ $F160W$ image was added to the redshifted galaxy images to simulate the effects of sky noise.
(Note that we do not conserve the luminosities of the local galaxy sample. Many
local galaxies would not be visible at $z=2-3$ in the rest-frame $u$ or $B$, and
Lyman break galaxies are typically two magnitudes brighter than local galaxies.
Our simulations in \S 3 suggest that at $<S/N> > 2$, spatial resolution will dominate any
morphological biases.)

We find that the poor spatial resolution of $z=2-3$ galaxies is expected to significantly
bias their observed morphologies.  In Table 8, we give the simulated mean biases in $C$, $A$,  $S$, $G$,
and $M_{20}$  for early, mid, and late-type galaxies at $z=2$ and $z=3$ observed at $\sim$ 1.3$\mu$m. 
A scatter of $\sim 0.13$ is introduced to the $A$ measurements, 
making it ineffective at distinguishing between early and late-type galaxies. 
$S$ also has large uncertainties at these resolutions, and large biases for E/S0s as a result
of their unresolved centers. $C$ and $M_{20}$ are also significantly biased as 
a function of morphological type, but have a greater dynamical range
and therefore are still useful. $G$ remains a reliable unbiased diagnostic out to at
least $z \sim 3$ for the NICMOS HDFN plate scale and PSF (Table 8).

Given these biases, the some of the observed LBG morphologies appear to be similar to the morphologies 
of local early-type galaxies (Figures 16-17).  However, some of the $z \sim 2-3$ galaxies have higher Gini coefficients
and/or asymmetries than expected from the degraded local galaxy images (Figure 16), and one $z \sim 3$
object has a double nucleus, resulting in a much higher asymmetry and $M_{20}$ than any of the degraded local galaxy
images.    We have applied a series of two-dimensional KS-tests
to the LBG sample and degraded local galaxy simulations, similar to the ones used in \S 4 (Table 9).   
We find that the $z \sim$ 2 LBG sample has a less than $0.4\%$ probability of matching
the degraded $B/g$ band local galaxy morphologies 
for all combinations of $C-A-G-M_{20}$, except for $C-M_{20}$ where 
systematic biases are the strongest.   The $z \sim 3$ LBGs are more likely to be drawn from a 
populations of galaxies with ``normal'' morphologies ($>$ 2\% probability); however fewer galaxies are
observed in the $z \sim 3$ and the $u$-band local galaxy samples, and one galaxy is highly asymmetric.
Therefore, it is highly unlikely that the $z \sim 2$ galaxies have morphologies identical
to local elliptical/S0 or spiral galaxies; rather their high $G$ and moderate $A$ values suggest that they are
more like the ULIRG population (Figures 10-12).

\section{SUMMARY}

We have re-defined the Gini coefficient in Equation 6 as a statistic for measuring the distribution of
flux values within a galaxy's image, and introduced $M_{20}$ (Equation 8), the second-order moment
of the brightest 20\% of the galaxy's flux.  These two indices are complementary, 
non-parametric morphology measures.  We have tested robustness of $G$ and $M_{20}$ 
to decreasing $S/N$ and resolution and found them to change by less than 10\% for average
$S/N$ per pixel $\geq$ 2 and resolutions better than 500 pc.   
At worse resolutions, $C$, $A$, and $M_{20}$ have 
systematic biases which are a function of Hubble type, while $S$ becomes unreliable. 
$G$, on the other hand, appears to be remarkably stable at low resolutions and
therefore is a powerful tool for classifying the morphologies of high-redshift galaxies.

We have measured $C$, $A$, $S$, $G$, and $M_{20}$ from the 
near-UV/optical images of 170 local E-S0-Sa-Sb-Sc-Sd-dI galaxies, 73 $z\sim 0.2$ ULIRGs,
and 49 $1.7 < z < 3.8$ Lyman break galaxies.  We find that:

1)Normal Hubble-type galaxies follow a tight $G-M_{20}-C$ sequence.  Early-type and
bulge-dominated systems have high Gini coefficients and concentrations 
and low second-order moments as a result of
their bright and compact bulges. Shallower surface brightness profiles, spiral
arms, and off-center star clusters give late-type disks lower Gini coefficients and 
concentrations and higher second-order moments. 

2) In combination with $A$ and $S$, $G$ is more effective than $C$ 
at distinguishing ULIRGs from normal Hubble types.  
We also find that most ULIRGs lie above the $G-M_{20}$ sequence and
can be identified by by their higher $G$ and $M_{20}$ values.
The high Gini coefficients arise from very bright compact nuclei, 
while multiple nuclei and bright tidal features produce large second-order moments.

3) ULIRGs with double and multiple nuclei have a statistically different distribution in morphology
space than single nuclei ULIRGs.  ULIRGs with double/multiple nuclei typically have
higher second-order moments and asymmetries and slightly lower concentrations than single
nuclei ULIRGs.  Singly-nucleated ULIRGs are more likely to possess low asymmetries and low
second-order moments, and often have higher concentrations and Gini coefficients than
ULIRGs in well-separated galaxy pairs.

4) Many of HDFN galaxies at $z \sim 2$ have higher rest-frame $B$-band Gini
coefficients and asymmetries than expected for local elliptical and spiral galaxies 
degraded to the same resolution. Instead, these objects are most 
similar in morphology to local ULIRGs.  

Our revised Gini coefficient has proven itself to be a highly robust and unbiased
non-parametric morphological indicator for $z > 2$ galaxies observed at HST NICMOS resolution,
and therefore has opened a window into the morphologies and assembly of the earliest galaxies. At
lower redshifts, and in combination with $M_{20}$, $A$, and $S$, the Gini coefficient allows us to
more precisely classify galaxy morphologies and identify merger candidates.  
In our next paper, we analyze a suite of hydrodynamical galaxy merger simulations 
to predict the evolution of merging galaxies in G-M-C-A-S morphology space.  
These simulations will explore a range of merger mass ratios, orbital parameters, and star-formation
feedback efficiencies, and will trace the spatial distribution of 
dark matter, gas, and old and new stars as a function of time (Cox et al. 2004). 

We would like to thank T.J. Cox and P. Jonsson for their valuable input and careful
reading of this manuscript.  We also gratefully acknowledge C. Conselice for his comments
and access to his morphology analysis code, M. Dickinson for use of 
the NICMOS Hubble Deep Field North observations, and the anonymous referee for their useful
suggestions. 
Support for J.L. was provided by NASA through grant number 9515  from the 
Space Telescope Science Institute, which is operated by AURA, Inc., under NASA contract NAS 5-26555.
J.P. acknowledges support from NSF through grant AST-0205944 and
NASA through NAG5-12326. P.M. acknowledges support by NASA through grants NAG5-11513
and GO-09425.19A from the Space Telescope Science Institute.

This work is based on observations made with the NASA/ESA Hubble Space Telescope, obtained from the data archive
at the Space Telescope Science Institute, and observations from the Sloan Digital Sky Survey.
Funding for the Sloan Digital Sky Survey (SDSS) has been provided by the Alfred P. Sloan Foundation, 
the Participating Institutions, the National Aeronautics and Space Administration, 
the National Science Foundation, the U.S. Department of Energy, the Japanese Monbukagakusho, 
and the Max Planck Society. 
The SDSS is managed by the Astrophysical Research Consortium (ARC) for the Participating Institutions. 
The Participating Institutions are The University of Chicago, Fermilab, the Institute for Advanced Study, 
the Japan Participation Group, The Johns Hopkins University, Los Alamos National Laboratory, 
the Max-Planck-Institute for Astronomy (MPIA), the Max-Planck-Institute for Astrophysics (MPA), 
New Mexico State University, University of Pittsburgh, Princeton University, 
the United States Naval Observatory, and the University of Washington.

\clearpage

\begin{deluxetable}{llrrrrrrrrl}
\tablecolumns{11}
\tabletypesize{\scriptsize}
\tablecaption{Test Galaxies}
\tablehead{ \colhead{Galaxy}  &\colhead{Type \tablenotemark{1}}  &\colhead{Dist} &\colhead{$S/N$} &\colhead{Res} 
&\colhead{$C_R$} &\colhead{$A_R$} &\colhead{$S_R$} &\colhead{$G_R$} &\colhead{$M20_R$} &\colhead{notes} \\
 &\colhead{} &\colhead{(Mpc)} &\colhead{} &\colhead{(pc/pix)} &\colhead{}
&\colhead{} &\colhead{} &\colhead{} &\colhead{} &\colhead{}}
\startdata
NGC 5332 \tablenotemark{2}  & E4(E/S0)     	&28.7  & 7.1  & 56   & 4.87  &-0.01  &-0.03  &0.63 & -2.66   & $-$	\\
NGC 4526 \tablenotemark{3}  & S0\_3\_(6)   	&17.0  &15.1  & 111  & 4.28  & 0.04  & 0.05  &0.59 & -2.40   & Virgo Cluster \\
NGC 3368 \tablenotemark{3}  & Sab(s)II      	&11.2  &14.6  & 73   & 3.98  & 0.06  & 0.06  &0.54 & -2.28   & Leo Group	 \\
NGC 3953 \tablenotemark{3}  & SBbc(r)I-II  	&18.6  & 8.4  & 122  & 3.54  & 0.08  & 0.20  &0.51 & -2.19   & Ursa Major Group \\
NGC 2403 \tablenotemark{3}  & Sc(s)III     	&3.2   & 5.8   & 19   & 3.02  & 0.07  & 0.34  &0.54 & -1.67   & M81 Group\\
NGC 4713 \tablenotemark{2}  & SAB(rs)d          &17.0  & 6.4  & 33   & 2.56  & 0.25  & 0.47   &0.47 & -1.52   & Virgo Cluster \\
Arp 220  \tablenotemark{4}  & ULIRG             &77.0  & 3.7  & 37  & 2.92  & 0.30  &0.43   &0.55 & -1.64   & IRAS15327$+$2340\\
SuperAntena \tablenotemark{4} & ULIRG          &245.4  & 3.1  &119  & 2.06  & 0.37  &1.04   &0.56 & -1.13   & IRAS19254$-$7245 \\
\enddata
\tablenotetext{1}{ Sandage \& Bedke 1994}
\tablenotetext{2}{ Abazajian et al. 2003}
\tablenotetext{3}{ Frei et al. 1996}
\tablenotetext{4}{ Borne et al. 2000}
\end{deluxetable}
\clearpage

\begin{deluxetable}{llrccrrrrrrrrrrr}
\setlength{\tabcolsep}{0.05in}
\tablewidth{0pc}
\tablecolumns{16}
\tabletypesize{\scriptsize}
\tablecaption{Frei Galaxy Catalog\tablenotemark{1}}
\tablehead{ \colhead{Galaxy} & \colhead{Type\tablenotemark{2}} & \colhead{M$_B$ \tablenotemark{3}}  & \colhead{($m-M$)} 
&\colhead{$S/N_B$}   &\colhead{C$_B$}  &\colhead{A$_B$} &\colhead{S$_B$}  &\colhead{G$_B$} &\colhead{M20$_B$} 
&\colhead{$S/N_R$}  &\colhead{C$_R$}  &\colhead{A$_R$}  &\colhead{S$_R$} &\colhead{G$_R$} &\colhead{M20$_R$} }
\startdata  
NGC 2768  &   S0\_1/2	     &  -20.9   & 31.75 \tablenotemark{6} &   7.7  &   4.33	     & -0.01  &    0.08 &     0.59  &	-2.39 &        7.9   &  4.32  & -0.02  &	  0.06   &   0.59  &   -2.45	  \\
NGC 3377  &   E6	     &  -19.0   & 30.25 \tablenotemark{6} &   5.6  &   4.77	     & -0.01  &   -0.01 &     0.63  &	-2.58 &        5.0   &  4.99  & -0.02  &	 -0.01   &   0.64  &   -2.67	  \\
NGC 3379  &   E1	     &  -19.9   & 30.12 \tablenotemark{6} &   9.0  &   4.61	     & -0.01  &   -0.02 &     0.59  &	-2.52 &        7.8   &  4.83  & -0.01  &	 -0.02   &   0.61  &   -2.54	  \\
NGC 4125{\tablenotemark{4}}  &  E6/S0\_1/2    &  -21.2   & 31.89 \tablenotemark{6} &   7.1  &   4.30   &  0.02  &  0.04 &   0.60  &-2.28 &     3.4    &  4.70 & -0.05  &	 0.00	&   0.63  &   -2.38 \\ 
NGC 4365  &   E3	     &  -21.0   & 31.55 \tablenotemark{6} &  3.4   &   4.40	     & -0.08  &   -0.06 &     0.61  &	-2.43 &        7.6   &  4.53  & -0.01  &	  0.01   &   0.59  &   -2.50	 \\
NGC 4374  &   E1	     &  -21.2   & 31.32 \tablenotemark{6} &  5.0   &   4.63	     & -0.06  &   -0.08 &     0.61  &	-2.46 &        7.6   &  4.57  & -0.05  &	 -0.05   &   0.60  &   -2.46	 \\
NGC 4472  &   E1/S0\_1\_(1)  &  -21.7   & 31.06 \tablenotemark{6} &  9.5  &	4.22	    & -0.02  &    0.00 &     0.58  &   -2.37 &        10.1   &  4.19  & -0.01  &   -0.01   &	0.58  &   -2.35  \\
NGC 4564  &   E6	     &  -19.0   & 30.33 \tablenotemark{5} &  6.5  &	4.81	    & -0.03  &    0.03 &     0.60  &   -2.44 &        12.5   &  5.29  &  0.01  &	 0.02	&   0.60  &   -2.45	 \\
\enddata				 				    		    									
\tablenotetext{1}{Frei et al. 1996, see also http://www.astro.princton.edu/\~frei/catalog.html }
\tablenotetext{2}{Sandage \& Bedke 1994}
\tablenotetext{3}{$B$ from RC3}
\tablenotetext{4}{also observed by SDSS Data Release 1, see Table 3}
\tablenotetext{5}{Distance from redshift obtained from the NASA Extragalactic Database, assuming $H_0 = 70$ km s$^{-1}$ Mpc$^{-1}$.}
\tablenotetext{6}{Jensen et al. 2003}\tablenotetext{7}{Freedman et al. 2001}\tablenotetext{8}{Gavazzi et al. 1999}	
\tablenotetext{9}{Garcia et al. 1996}\tablenotetext{10}{Theureau et al. 1997}\tablenotetext{11}{Yasuda et al. 1997}
\tablenotetext{12}{Eastman, Schmidt, \& Kirshner 1996}	\tablenotetext{13}{Parodi et al. 2000}
\tablenotetext{14}{Russell 2002}\tablenotetext{15}{Shanks 1997}	\tablenotetext{16}{Drozdovsky \&  Karachentsev 2000}
\tablenotetext{17}{Schoniger \& Sofue 1994} \tablenotetext{18}{Pierce 1994}
\tablenotetext{19}{Pierce \& Tully 1988}\tablenotetext{20}{Pierce, McClure, Racine 1992}
\tablenotetext{21}{Karachentsev, Musella, Grimaldi 1996}
\tablenotetext{22}{Karachentsev et al. 2002}
\tablenotetext{23}{Karachentsev \& Drozdovsky 1998}
\end{deluxetable}

\clearpage

\begin{deluxetable}{llrccrrrrrrrrrrrrrrrrrr}
\setlength{\tabcolsep}{0.05in}
\tablecolumns{22}
\tabletypesize{\scriptsize}
\rotate
\tablewidth{0pc}
\tablecaption{SDSS u-selected Catalog \tablenotemark{1}}
\tablehead{ \colhead{Galaxy} & \colhead{Type \tablenotemark{2}}  & \colhead{M$_B$ \tablenotemark{3}} &\colhead{($m-M$)}
&\colhead{S/N$_u$}   &\colhead{C$_u$}  &\colhead{A$_u$} &\colhead{S$_u$} &\colhead{G$_u$} &\colhead{M20$_u$} 
&\colhead{S/N$_g$}  &\colhead{C$_g$}  &\colhead{A$_g$}  &\colhead{S$_g$} &\colhead{G$_g$} &\colhead{M20$_g$} 
&\colhead{S/N$_r$}  &\colhead{C$_r$}  &\colhead{A$_r$}  &\colhead{S$_r$} &\colhead{G$_r$} &\colhead{M20$_r$}}
\startdata  
NGC 3640  & E3                            &  -19.0  &  30.37 \tablenotemark{6}  &  1.6   &  4.26	&    -0.14   &    -0.13   &   0.60   &    -2.34   &  6.1   &  4.55   &    0.00  &	0.00	&    0.63    &   -2.45   &  8.3    &   4.45  &   0.01   &   0.00   & 0.62   &  -2.45 \\  
NGC 4073  & E5                            &  -22.2  &  34.65 \tablenotemark{5}  &  0.7   &  3.95	&    -0.22   &    -0.04   &   0.47   &    -2.02   &  2.1   &  4.51   &   -0.08  &    -0.02	&    0.62    &   -2.58   &  2.8    &   4.64  &  -0.06   &  -0.03   & 0.62   &  -2.58 \\  
NGC 4125\tablenotemark{4} & E6/S0\_1/2    &  -21.2  &  31.89 \tablenotemark{6}  &  0.9   &  4.37	&    -0.17   &    -0.11   &   0.56   &    -2.36   &  5.2   &  4.54   &   -0.04  &	0.07	&    0.63    &   -2.51   &  7.1    &   4.64  &  -0.04   &   0.07   & 0.63   &  -2.54 \\  
NGC 4261  & E3                            &  -21.1  &  32.50 \tablenotemark{6}  &  0.9   &  4.56	&    -0.07   &     0.48   &   0.55   &    -2.55   &  4.4   &  4.64   &   -0.03  &    -0.08	&    0.62    &   -2.59   &  6.8    &   4.71  &  -0.01   &  -0.05   & 0.62   &  -2.58 \\  
NGC 4636\tablenotemark{4}  & E0/S0\_1\_(6)&  -20.4  &  30.83 \tablenotemark{6}  &  0.6  &  3.88    &    -0.23   &	-0.04	&   0.46   &	-1.63	  &  2.9   &  4.28   &   -0.06  &    -0.10	&    0.60    &   -2.54   &  3.9    &   4.33  &  -0.05   &  -0.08   & 0.60   &  -2.56 \\    
NGC 5322\tablenotemark{4}  & E4(E/S0)     &  -21.3  &  32.47 \tablenotemark{6}  &  1.0  &  4.70    &    -0.07   &	 0.72	&   0.57   &	-2.58	  &  5.4   &  4.84   &   -0.01  &    -0.01	&    0.63    &   -2.66   &  7.1    &   4.87  &  -0.01   &  -0.03   & 0.63   &  -2.66 \\  
NGC 5813\tablenotemark{4}  & E1  	  &  -21.1  &  32.54 \tablenotemark{6}  &  0.7  &  4.14    &    -0.30   &	-0.76	&   0.48   &	-2.26	  &  3.3   &  4.38   &   -0.05  &    -0.10	&    0.60    &   -2.57   &  4.3    &   4.48  &  -0.04   &  -0.08   & 0.60   &  -2.60 \\
\enddata	
\tablenotetext{1}{Abazajian et al. 2003; http://www.sdss.org/dr1}
\tablenotetext{2}{Sandage \& Bedke 1994}
\tablenotetext{3}{$B$ from RC3}
\tablenotetext{4}{also observed by Frei et al. 1996; see Table 2}
\tablenotetext{5}{Distance from redshift obtained from the NASA Extragalactic Database, assuming $H_0 = 70$ km s$^{-1}$ Mpc$^{-1}$.}
\tablenotetext{6}{Jensen et al. 2003}
\tablenotetext{7}{Freedman et al. 2001}	
\tablenotetext{8}{Gavazzi et al. 1999}	
\tablenotetext{9}{Shanks 1997}	
\end{deluxetable}																			   

\begin{deluxetable}{llrrrrrrrr}
\setlength{\tabcolsep}{0.05in}
\tablecolumns{10}
\tabletypesize{\scriptsize}
\tablewidth{0pc}
\tablecaption{Van Zee Dwarf Irregular Galaxy Sample \tablenotemark{1}}
\tablehead{ \colhead{Galaxy} & \colhead{Type \tablenotemark{2}} &\colhead{$M_B$}
&\colhead{S/N$_B$}   &\colhead{C$_B$}  &\colhead{A$_B$} &\colhead{S$_B$} &\colhead{G$_B$} &\colhead{M20$_B$}} 
\startdata  
UGC00290   &  Sdm     & -14.48  &  2.1  &  3.75   &    0.03 &	0.30  &  0.54  &    -1.78 \\
UGC00634   &  SABm    & -17.67  &  2.5  &  2.85   &    0.14 &	0.61  &  0.49  &    -1.15 \\
UGC00685   &  SAm     & -14.74  &  5.0  &  2.81   &    0.18 &   0.39  &  0.52  &    -1.47 \\  
UGC00891   &  SABm    & -15.53  &  3.1  &  2.90   &    0.06 &   0.38  &  0.49  &    -1.56 \\  
UGC01104   &  Im      & -16.08  &  9.2  &  3.18   &    0.20 &   0.36  &  0.50  &    -1.69 \\  
UGC01175   &  Sm      & -14.13  &  3.8  &  2.84   &    0.24 &   0.27  &  0.47  &    -1.38 \\  
UGC03647   &  IBm     & -17.06  &  2.0  &  2.87   &    0.12 &	0.56  &  0.54  &    -1.47 \\  
UGC04117   &  IBm     & -14.86  &  4.7  &  2.89   &    0.28 &   0.67  &  0.48  &    -0.94 \\
\enddata
\tablenotetext{1}{Van Zee 2001}
\tablenotetext{2}{Morphological types from RC3}
\end{deluxetable}

\clearpage	

\begin{deluxetable}{llrrrrrrrrl}
\tablecolumns{11}
\tabletypesize{\scriptsize}
\tablecaption{Borne et al. 2000 ULIRG HST Survey\tablenotemark{1}}
\tablehead{ \colhead{Galaxy} & \colhead{redshift \tablenotemark{2}} & \colhead{RA} &\colhead{DEC} & \colhead{S/N} 
 &\colhead{$C$} &\colhead{$A$} &\colhead{$S$} &\colhead{$G$} &\colhead{$M_{20}$} &\colhead{Nuclei \tablenotemark{3}} }
\startdata
IRAS00091$-$0738     &      0.118    &   00 11 43  &   $-$07 22 07    &   2.0   &   3.63  &	0.20	&     0.38  &    0.59  &  $-$1.91   &    Single$^*$\\
IRAS00104$-$0139     &      0.163    &   00 13 04  &   $-$01 23 05    &   2.1   &   2.20  &	0.26	&     0.47  &    0.52  &  $-$1.22   &    Double     \\ 
IRAS00161$-$0850     &      0.109    &   00 18 43  &   $-$08 33 36    &   3.7   &   2.69  &	0.28	&     0.54  &    0.56  &  $-$1.26   &    Multiple \\ 
IRAS00207+1029       &      0.231    &   00 23 22  &   $+$10 46 22    &   3.3   &   3.79  &	0.52	&     0.90  &    0.73  &  $-$1.74   &    Double     \\
IRAS00335$-$2732     &      0.069    &   00 36 01  &   $-$27 15 35    &   5.0   &   4.33  &	0.15	&     0.68  &    0.65  &  $-$2.11   &    Single     \\ 
IRAS00456$-$2904     &      0.11     &   00 48 07  &   $-$28 48 19    &   3.9   &   3.31  &	0.21	&     1.59  &    0.60  &  $-$1.66   &    Double$^*$	 \\
IRAS01031$-$2255     &      0.187    &   01 05 37  &   $-$22 39 18    &   2.0   &   4.12  &	0.23	&     0.13  &    0.40  &  $-$1.00   &    Single     \\
IRAS01199$-$2307     &      0.156    &   01 22 21  &   $-$22 52 07    &   2.2   &   4.72  &	0.22	&     0.27  &    0.51  &  $-$0.89   &    Double$^*$	 \\
\enddata								     
\tablenotetext{1}{Borne et al. 2000; HST WFPC2 F814W observations}
\tablenotetext{2}{redshifts taken from NASA Extragalactic Database}
\tablenotetext{3}{Cui et al. 2002 classification, except where starred.}
\end{deluxetable}

\clearpage
\begin{deluxetable}{cllll}
\tablecolumns{5}
\tablewidth{0pc}
\tabletypesize{\scriptsize}
\tablecaption{Two-Dimensional KS Test Probabilities}
\tablehead{ \colhead{}  &\colhead{} &\colhead{} &\colhead{} &\colhead{} }
\startdata
\cutinhead{``Normal'' Galaxies v. ULIRGs }
       &  $A$      &  $S $      & $G $       & $M_{20}$ \\
$C$      & 6.6e-15 & 9.3e-8  & 7.5e-7    & 8.3e-8	 \\
$A$      &  ...    & 2.1e-16 & 7.7e-18   & 6.2e-14  \\
$S$      &  ...    & ...     & 1.6e-13   & 5.1e-10  \\
$G$      &  ...    & ...   & ...         & 5.8e-13  \\
\cutinhead{ Double v. Multi-Nuclei ULIRGs}
       &  $A$      &  $S$       & $G$        & $M_{20}$ \\
$C$      & 0.29    & 0.25    & 0.15    & 0.66	\\
$A$      &  ...    & 0.047   & 0.17    & 0.086       \\
$S$      &  ...    & ...     & 0.086   & 0.36         \\
$G$      &  ...    & ...     & ...     & 0.18   \\
\cutinhead{ Paired v. Single-Nuclei ULIRGs}
       &  $A$      &  $S$       & $G$        & $M_{20}$ \\
$C$       & 0.25    & 0.21   & 0.29   & 0.14	\\
$A$       &  ...    & 0.51   & 0.49   & 0.13         \\
$S$       &  ...    & ...    & 0.47   & 0.30        \\
$G$       &  ...    & ...   & ...     & 0.12   \\
\cutinhead{ Double+Multi- v. Single-Nuclei ULIRGs}
       &  $A$      &  $S$       & $G$        & $M_{20}$ \\
$C$       & 8.2e-7 & 0.0012  & 0.029   & 3.3e-4	\\
$A $      &  ...   & 7.1e-6  & 7.2e-6  & 9.3e-7      \\
$S$       &  ...    & ...    & 0.0024  & 3.2e-5         \\
$G$       &  ...    & ...     & ...    & 7.6e-4  \\
\cutinhead{ Double+Multi-Nuclei v. Paired ULIRGs}
       &  $A$      &  $S$       & $G$        & $M_{20}$ \\
$C$    & 0.0041  & 0.017   & 0.078    & 0.057	\\
$A$    &  ...    & 0.0042  & 0.020    & 0.0070         \\
$S$    &  ...    & ...    &  0.096    & 0.016         \\
$G$    &  ...    & ...   & ...        & 0.056  \\
\enddata
\tablecomments{Two-dimensional K-S test probabilities that
two galaxy populations have the same distribution in a two-parameter
space.}
\end{deluxetable}

\clearpage
\begin{deluxetable}{rlcccrrrrrrr}
\tablecolumns{12}
\tablewidth{0pc}
\tabletypesize{\scriptsize}
\tablecaption{HDFN Lyman Break Galaxies \tablenotemark{1}}
\tablehead{ \colhead{ID \tablenotemark{1}} & \colhead{redshift \tablenotemark{2}} 
&\colhead{RA (J2000)} &\colhead{DEC (J2000)} &\colhead{$H$\tablenotemark{1}} &\colhead{M$_B$}
&\colhead{r$_p$ (\arcsec)} &\colhead{$S/N$} &\colhead{$C$} &\colhead{$A$} &\colhead{$G$}
&\colhead{$M_{20}$}}
\startdata
  882  &  1.8 &  12:36:45.66  &  62:12:41.9 &  22.27&  -21.6  &   0.62 & 8.2  &  4.4  &   0.08  &  0.64   &  $-$1.86 \\ 
 1118  &  1.8 &  12:36:51.28  &  62:12:33.8 &  24.47&  -19.7  &   0.31 & 4.0  &  3.3  &   0.12  &  0.63   &  $-$1.45 \\ 
 1217  &  1.8 &  12:36:45.15  &  62:12:05.5 &  24.77&  -19.5  &   0.41 & 3.1  &  3.5  &   0.02  &  0.55   &  $-$1.19 \\ 
  121  &  1.9 &  12:36:44.85  &  62:14:06.1 &  23.14&  -20.9  &   0.36 & 4.2  &  4.0  &   0.07  &  0.68   &  $-$1.77  \\ 
  653  &  1.9 &  12:36:56.64  &  62:13:39.9 &  23.87&  -20.3  &   0.63 & 4.0  &  4.0  &   0.05  &  0.62   &  $-$1.52  \\ 
  769  &  1.9 &  12:36:52.00  &  62:13:14.9 &  24.86&  -19.5  &   0.48 & 3.5  &  3.5  &   0.05  &  0.62   &  $-$1.68 \\ 
 1550  &  1.9 &  12:36:58.29  &  62:12:16.5 &  23.75&  -20.3  &   0.48 & 3.4  &  4.2  &  -0.08  &  0.63   &  $-$1.68  \\ 
  539  &  2.0 &  12:36:50.29  &  62:13:29.8 &  23.74&  -20.5  &   0.39 & 3.5  &  4.1  &  -0.02  &  0.64   &  $-$1.91  \\
\enddata
\tablenotetext{1}{M. Dickinson, private communication.}
\tablenotetext{2}{Photometric redshifts; see Budavari et al. 2000}
\tablenotetext{3}{Spectroscopic redshifts; see Cohen et al. 2000}
\end{deluxetable}

\begin{deluxetable}{lrrrcrr}
\tablecolumns{7}
\tablewidth{0pc}
\tabletypesize{\scriptsize}
\tablecaption{z=2 and z=3 morphological simulations}
\tablehead{ \colhead{Type} & \colhead{N}  
&\colhead{$C_{z}- C_{z=0}$} &\colhead{$A_{z} - A_{z=0}$} &\colhead{$S_{z} - S_{z=0}$}   
&\colhead{$G_{z}- G_{z=0}$} &\colhead{$M20_{z} - M20_{z=0}$}}
\startdata
\cutinhead{z=2, rest-frame $B/g$ observations}
E/S0   & 36   & +0.14 $\pm$ 0.69 &  +0.07 $\pm$ 0.05  & +0.29 $\pm$ 0.19 & -0.01 $\pm$ 0.02 & +0.54 $\pm$ 0.26  \\
Sa-Sbc & 58   & +0.36 $\pm$ 0.51 &  +0.08 $\pm$ 0.15  & +0.05 $\pm$ 0.22 & -0.01 $\pm$ 0.03 & +0.23 $\pm$ 0.41 \\
Sc-Sd  & 58   & +0.46 $\pm$ 0.37 &  +0.03 $\pm$ 0.18  & -0.10 $\pm$ 0.28 & -0.01 $\pm$ 0.04 & -0.06 $\pm$ 0.20  \\
dI     & 22   & +0.07 $\pm$ 0.66 &  -0.07 $\pm$ 0.12  & +0.46 $\pm$ 0.53 & +0.10 $\pm$ 0.07 & -0.16 $\pm$ 0.28  \\
\cutinhead{z=3, rest-frame $u$ observations}
E/S0    & 3  & +0.04 $\pm$ 0.74 & +0.09 $\pm$ 0.23  & +0.56 $\pm$ 0.52  & -0.02 $\pm$ 0.10 & +0.95 $\pm$ 0.40 \\
Sa-Sbc  & 5  & +0.50 $\pm$ 0.72 & +0.11 $\pm$ 0.13  & -0.10 $\pm$ 0.31  & -0.02 $\pm$ 0.03 & -0.08 $\pm$ 0.49 \\
Sc-Sd   & 8  & +0.47 $\pm$ 0.58 & -0.02 $\pm$ 0.13  & -0.34 $\pm$ 0.40  & +0.02 $\pm$ 0.07 & -0.29 $\pm$ 0.30 \\
\enddata
\end{deluxetable}

\begin{deluxetable}{clllclll}
\tablecolumns{8}
\tablewidth{0pc}
\tabletypesize{\scriptsize}
\tablecaption{Lyman Break Galaxy KS Test Probabilities}
\tablehead{ 
\colhead{}  & \multicolumn{3}{c}{$z=2$ LBGs v. local galaxies } 
&\colhead{}  & \multicolumn{3}{c}{$z=3$ LBGs v. local galaxies } }
\startdata
       & $ A$       & $ G $     & $M_{20}$ &  & $ A$ &$ G$ &$M_{20}$ \\
$C$      & 5.1e-4   & 4.6e-6  & 0.075 &
         & 0.065     & 0.032  & 0.27	 \\
$A$      &  ...     & 2.1e-8  & 0.0039 &
       &  ...       & 0.017   & 0.13       \\
$G$      &  ...     & ...     & 6.6e-6  &  
         &  ...     & ...     & 0.10     \\
\enddata
\tablecomments{Two-dimensional K-S test probabilities that
two galaxy populations have the same distribution in a two-parameter
space.}
\end{deluxetable}

\clearpage
\begin{figure}
\epsscale{0.5}
\plotone{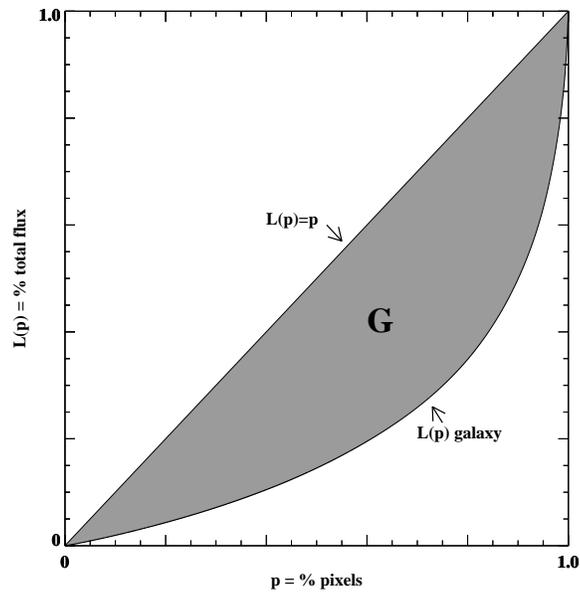}
\caption{ The Lorenz curve -- the Gini coefficient is the area between the Lorenz curve of the galaxy's pixels, 
and that of equitable distribution (shaded region). The given curve is for S0 NGC4526, G=0.59}
\end{figure}

\clearpage
\begin{figure}
\epsscale{1.0}
\plottwo{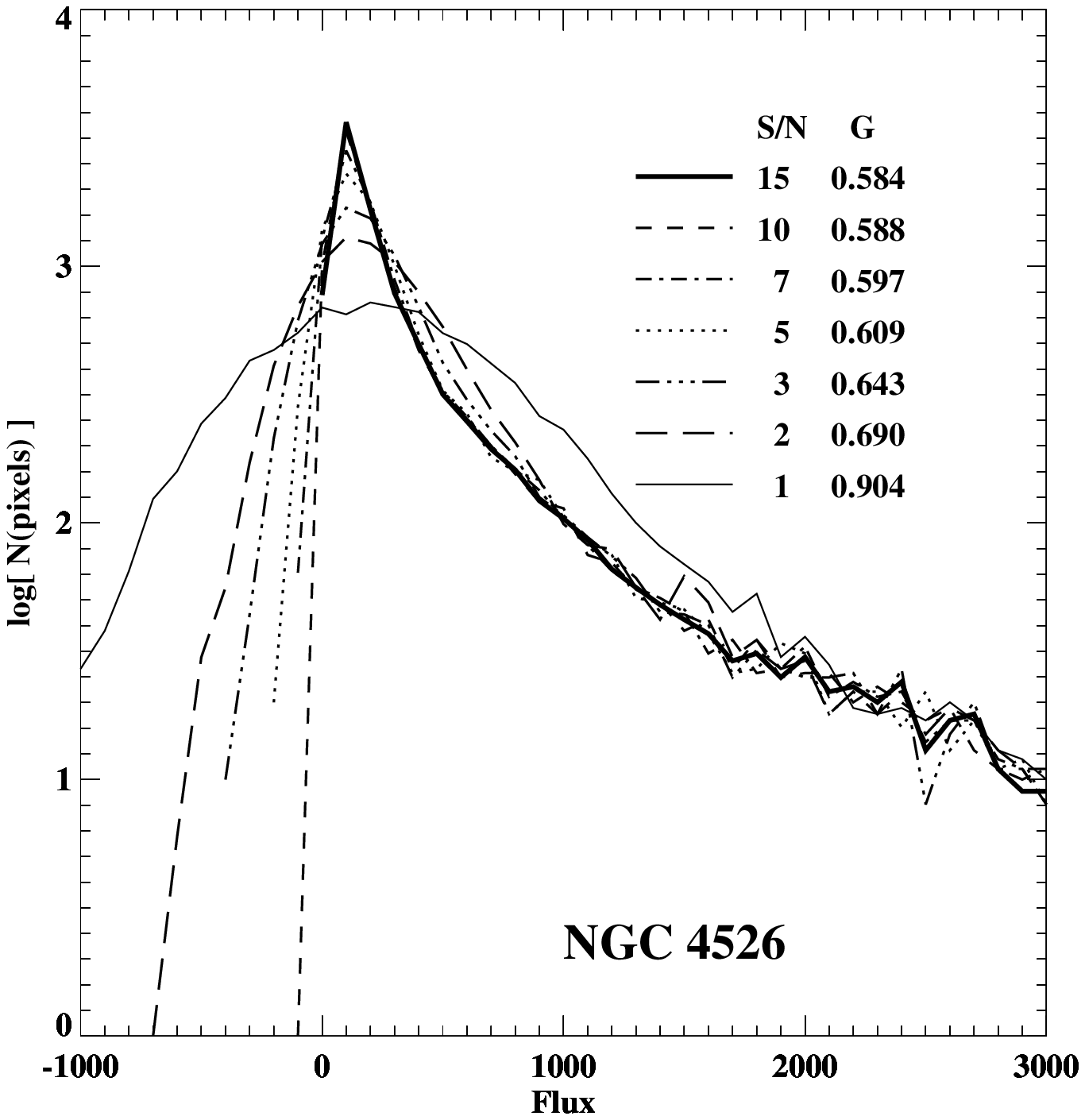}{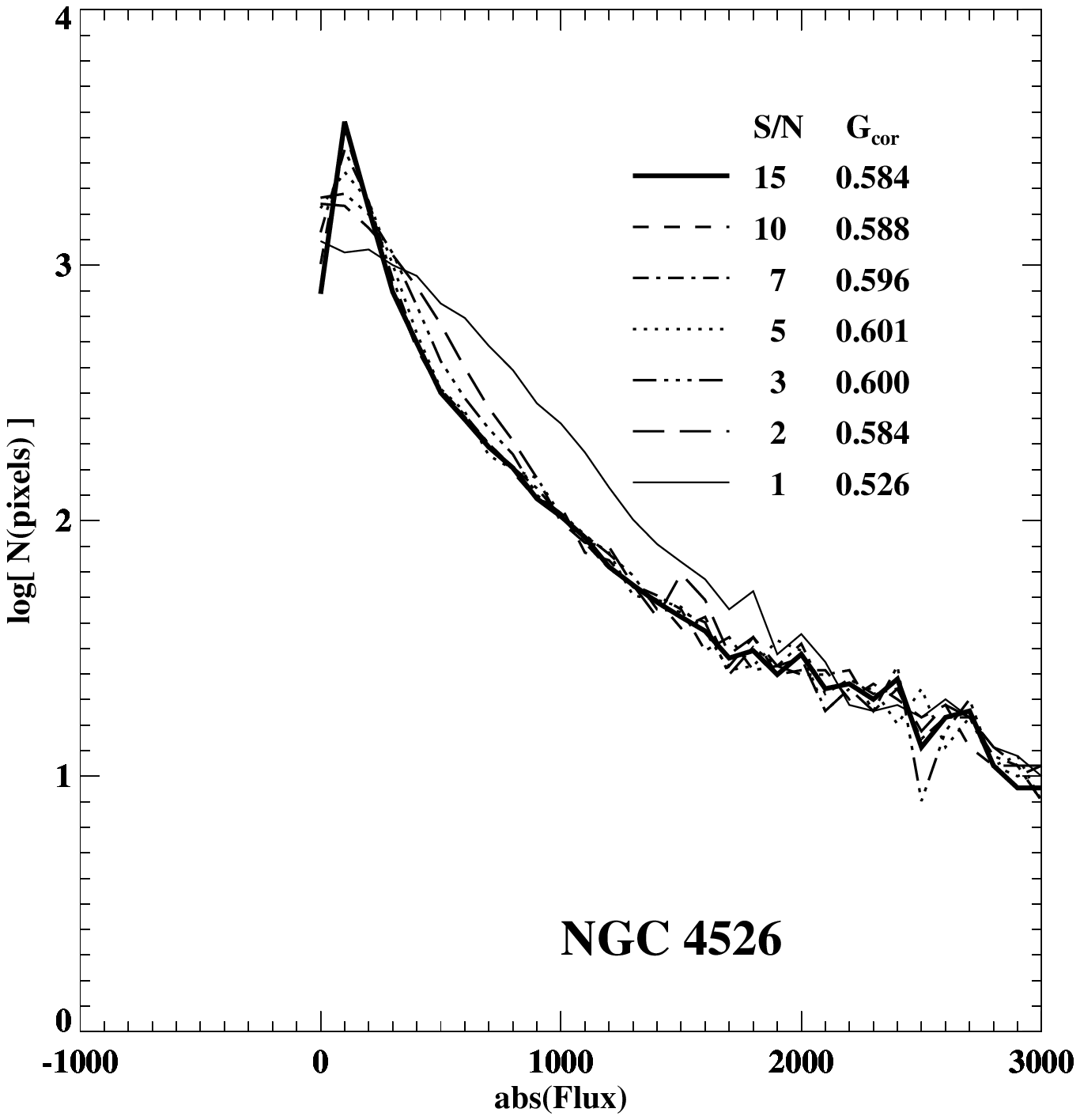}
\caption{ The pixel flux value distribution as a function of the average signal-to-noise per galaxy pixel
for S0 galaxy NGC4526. Left: As $<S/N>$ decreases, more faint galaxy pixels are scattered below the background sky
level. Right: The corrected Gini coefficients are calculated from the distribution of absolute pixel flux
values.}
\end{figure}

\clearpage
\begin{figure}
\epsscale{0.8}
\caption{ [Lotz.fig3.gif]  Test galaxy morphological measurements $C$, $A$, $S$, $G$, and $M_{20}$ for rest-frame $\sim 6500$\AA\
images (Table 1).  In the first panel, inner and outer circles enclose 20\% and 80\% of the flux within 1.5 $r_p$. 
The second panel shows the residual $I- I_{180}$ image, with the circle at 1.5 $r_p$. 
The third panel shows the residual $I- I_{S}$ image, with the inner and outer circles at 0.25 and 1.5 $r_p$.
The fourth panel images are the original galaxy images scaled such that
the minimum surface brightness matches that used to create the galaxy segmentation maps.  The outer edge of
the segmentation map are the outer contour plotted in the fourth and fifth panels. 
The inner contours plotted in the fifth panel trace each galaxy's brightest 20\% of it flux, while the crosses 
indicate each galaxy's center. The final panel plots each galaxy's $G$ and $M_{20}$ where the solid line is for reference. }
\end{figure}

\begin{figure}
\epsscale{0.8}
\caption{[Lotz.fig4.gif]  Same as Figure 3 }
\end{figure}

\clearpage
\begin{figure}
\epsscale{1.0}
\plottwo{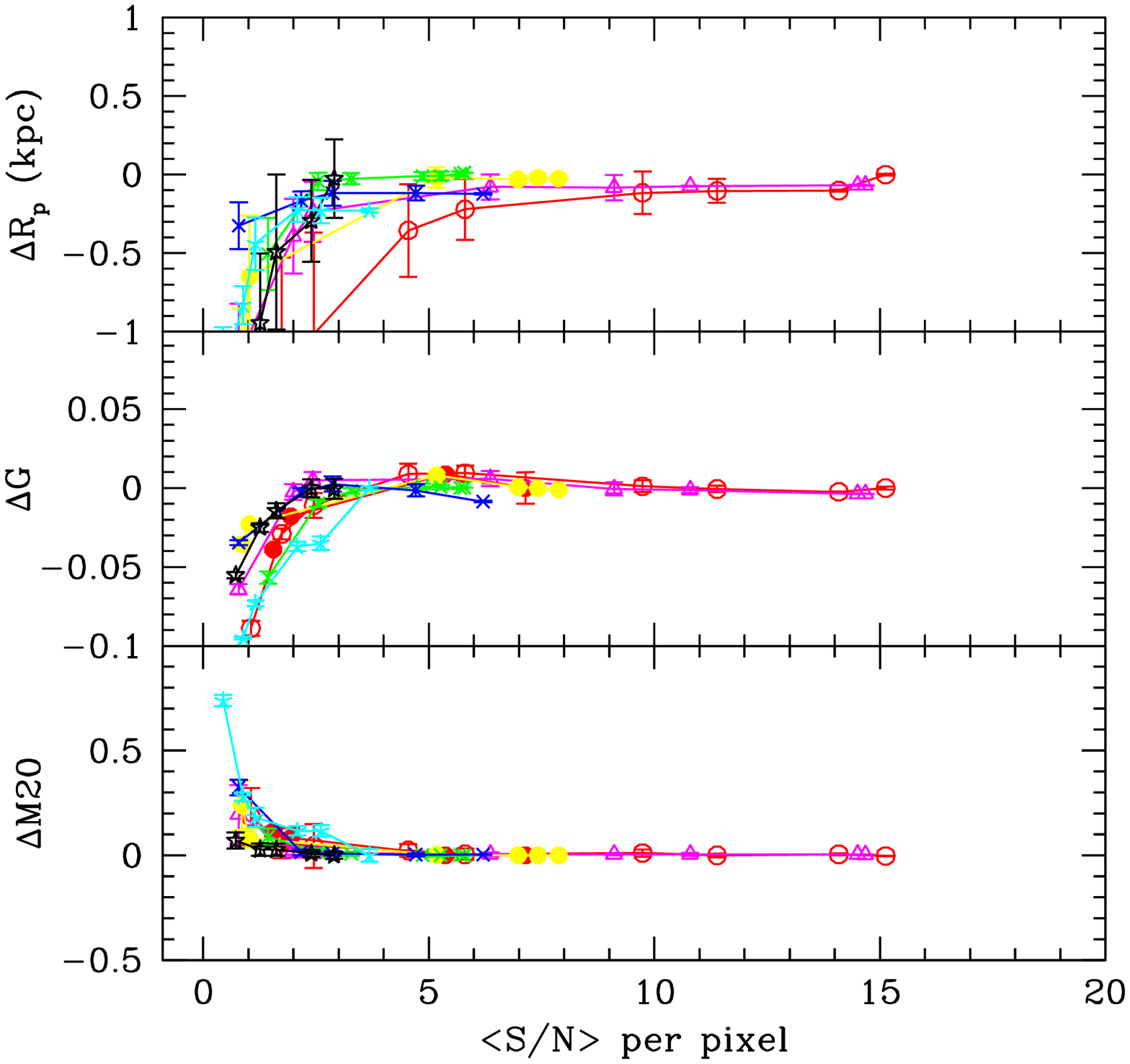}{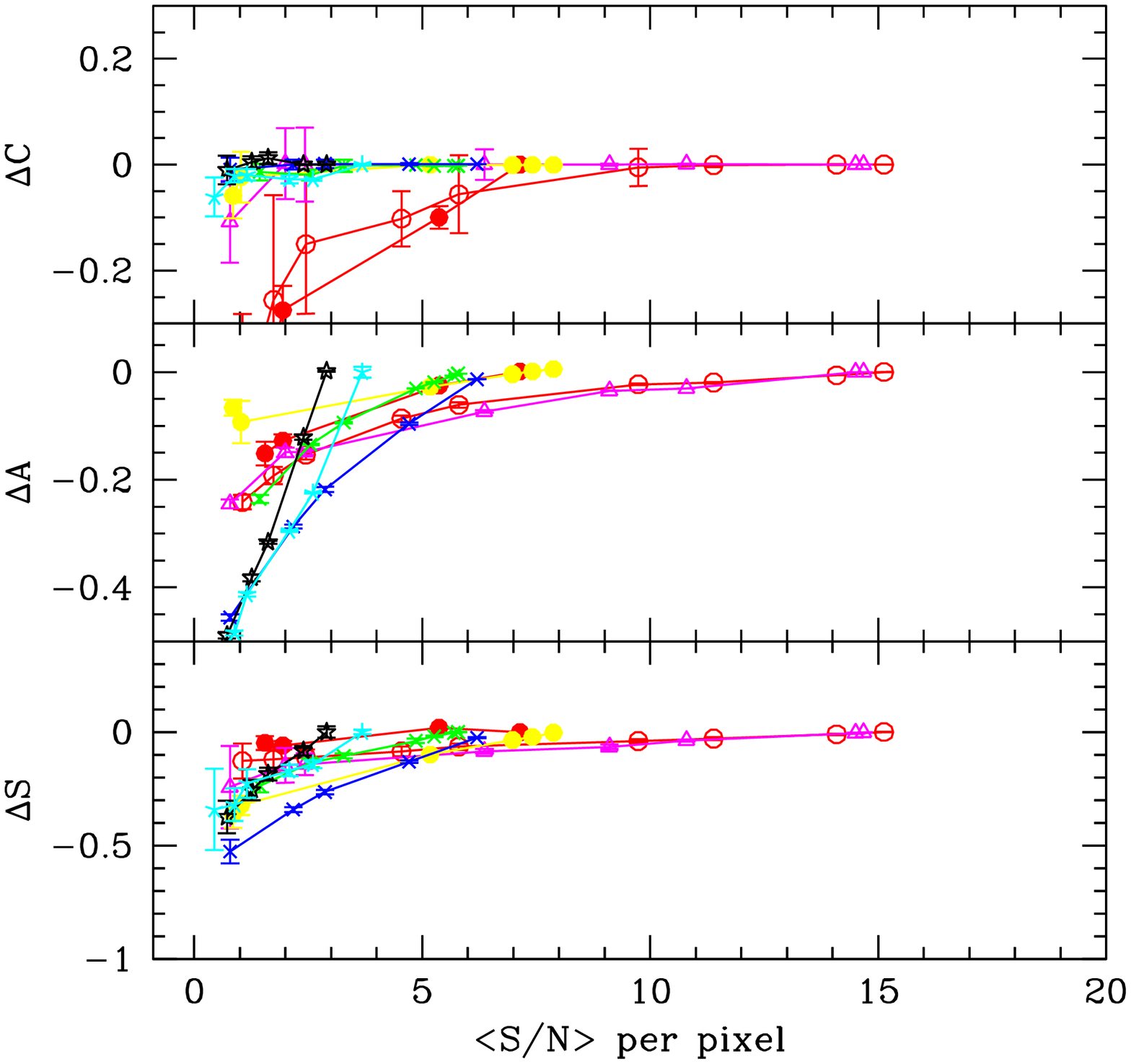}
\caption{ $\Delta r_p$, $\Delta$G, $\Delta M_{20}$, $\Delta$C, $\Delta$A, and $\Delta$S  vs. $<S/N>$ per pixel. 
E/NGC5322 = red filled circles, S0/NGC4526= red open circles, Sab/NGC3368= magenta triangles, Sbc/NGC3953= yellow
squares, Sc/NGC2403=green crosses, Sd/NGC4713=dark blue crosses, Arp220=light blue stars, SuperAntena=black stars }
\end{figure}

\clearpage
\begin{figure}
\plottwo{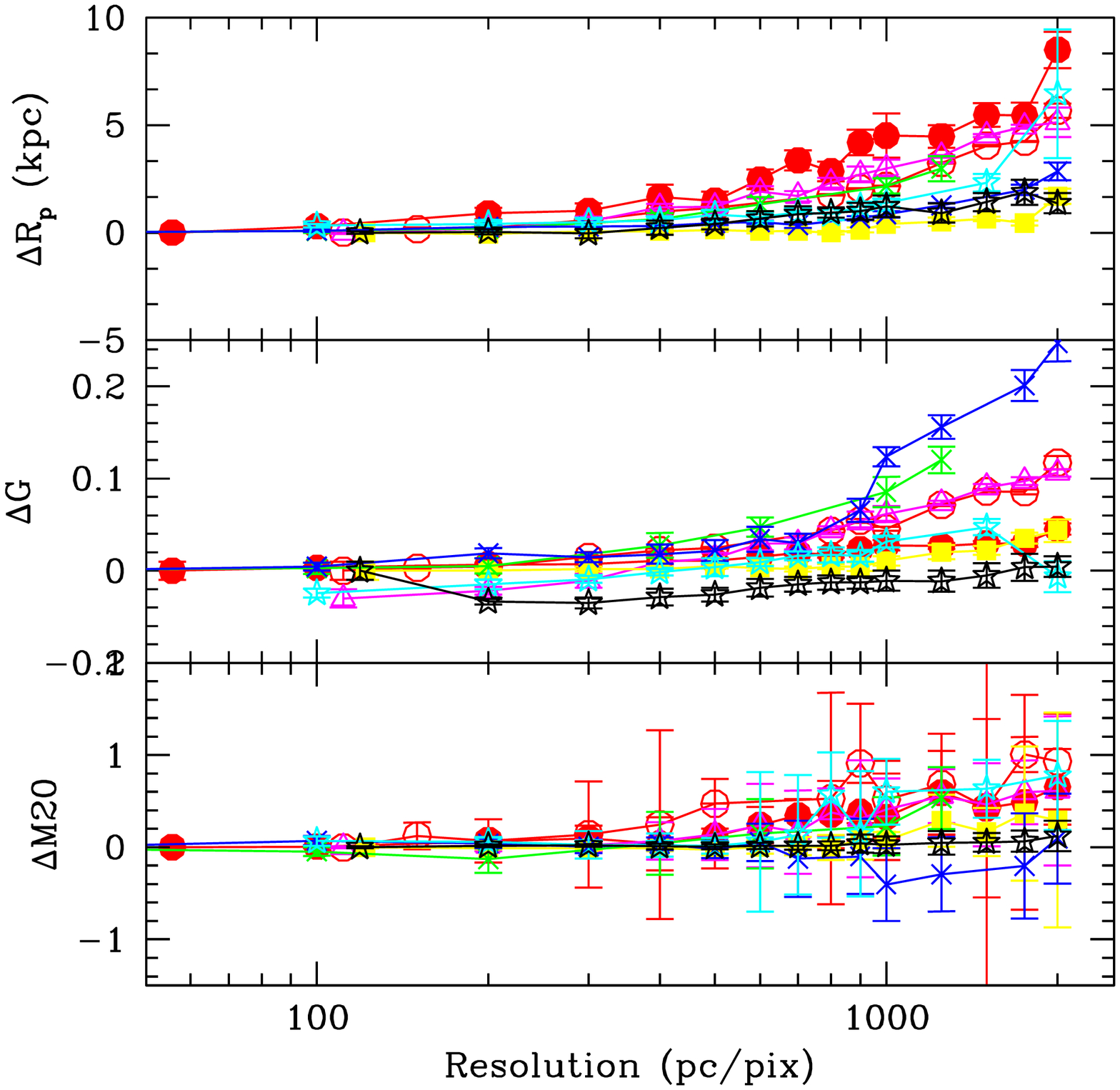}{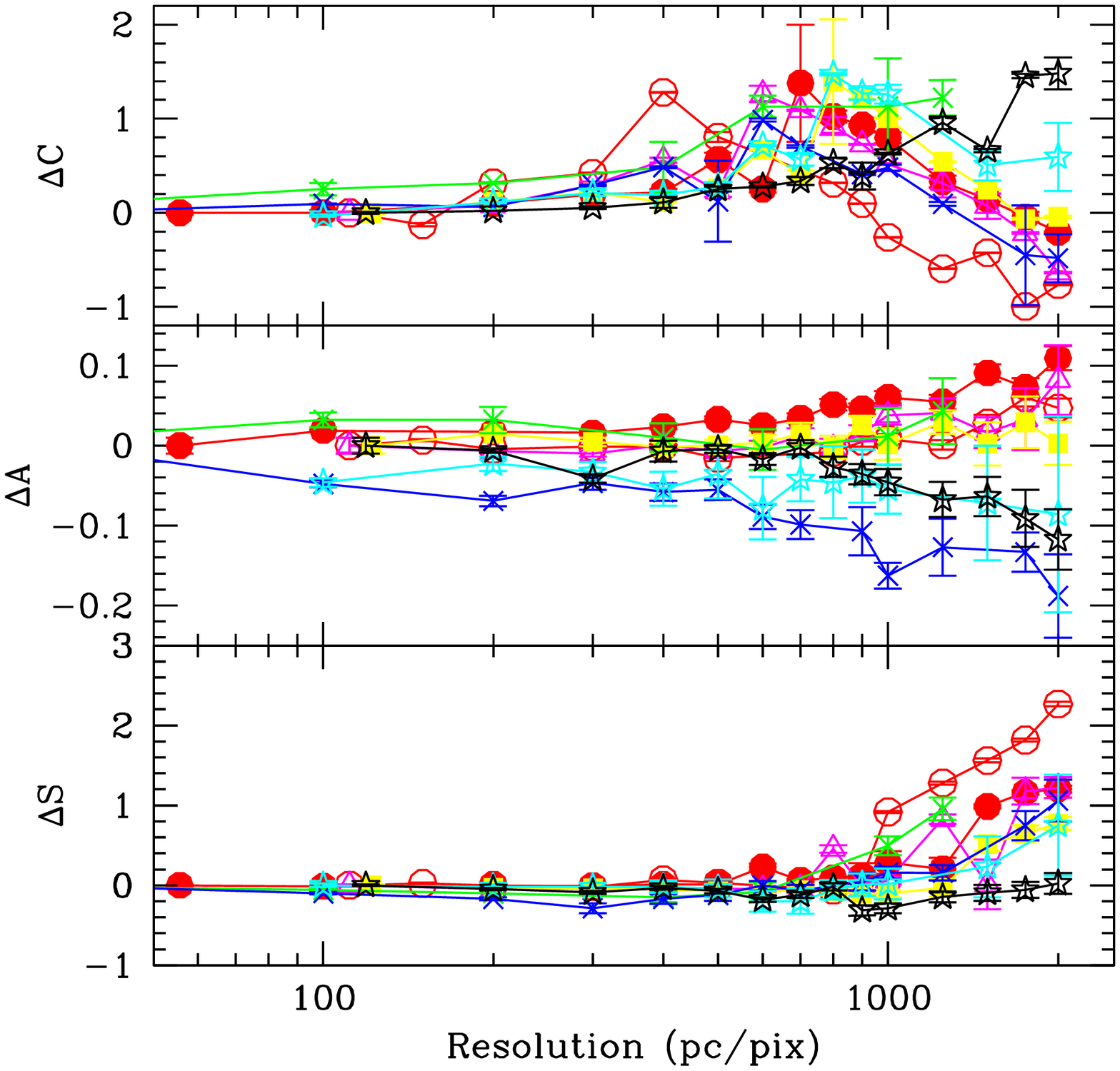}
\caption{ $\Delta r_p$, $\Delta$G,  $\Delta$M$_{20}$, $\Delta$C, $\Delta$A, and $\Delta$S vs. resolution (pc per pixel).
Symbols are same as Figure 5.  \label{sims_res}}
\end{figure}

\clearpage
\begin{figure}
\epsscale{1.0}
\plotone{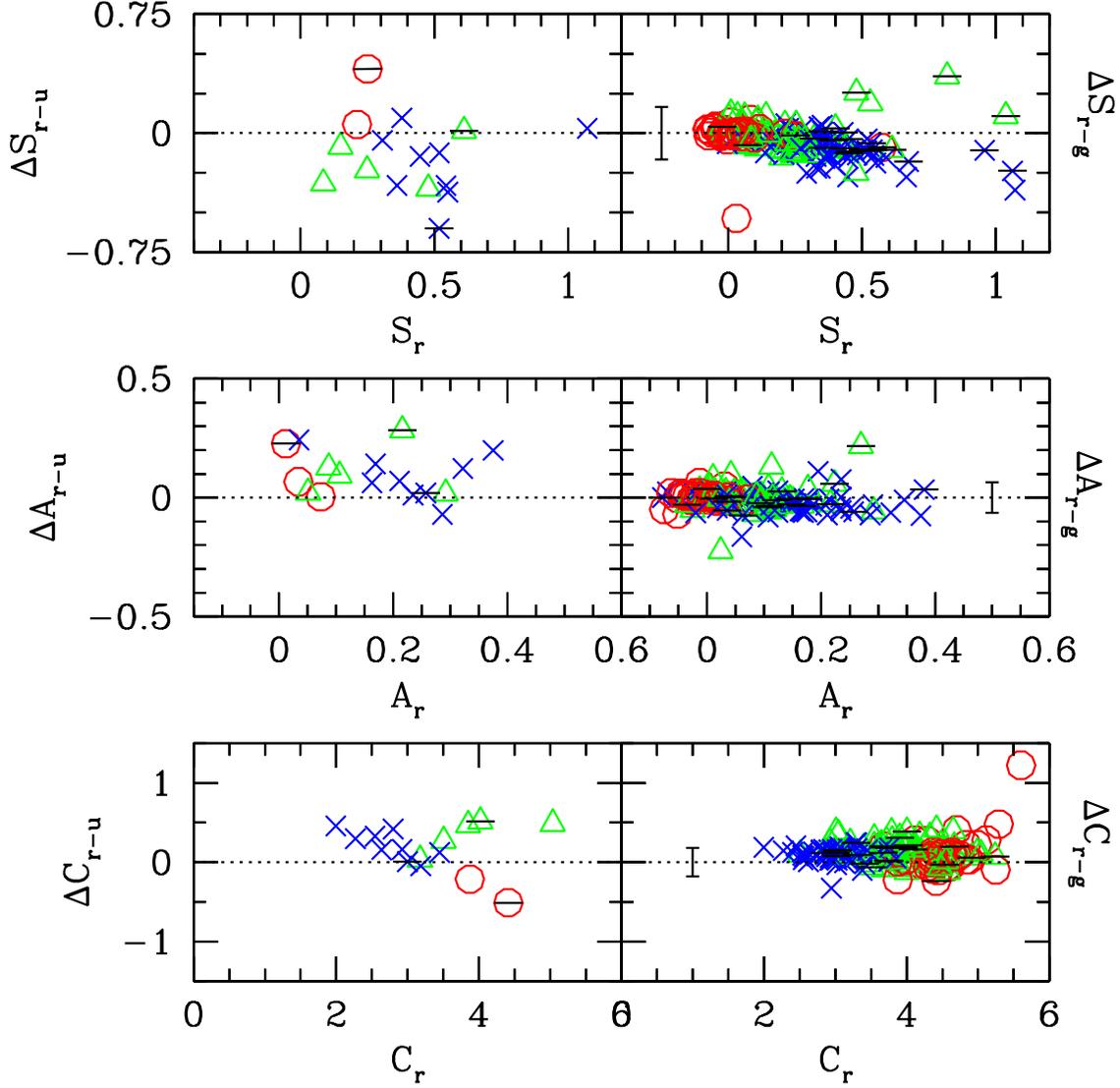}
\caption{ Change in $CAS$ morphology from $\sim$ 6500\AA\ ($R/r$) to $\sim$ 3600\AA\ ($u$) 
for SDSS u-selected sample with $S/N_u > 2.0$ (left)
and to $\sim$ 4500\AA\ ($B/g$) for Frei and SDSS galaxies with $S/N > 2.0$ (right). The error-bars
are $\sqrt{(\delta_r^2 + \delta_g^2)}$, where $\delta$ is the average difference between SDSS and
Frei et al. observations of the same galaxies. E/S0 are red circles, Sa-Sbc are green triangles, and Sc-Sdm are
blue crosses.}
\end{figure}

\clearpage
\begin{figure}
\epsscale{1.0}
\plotone{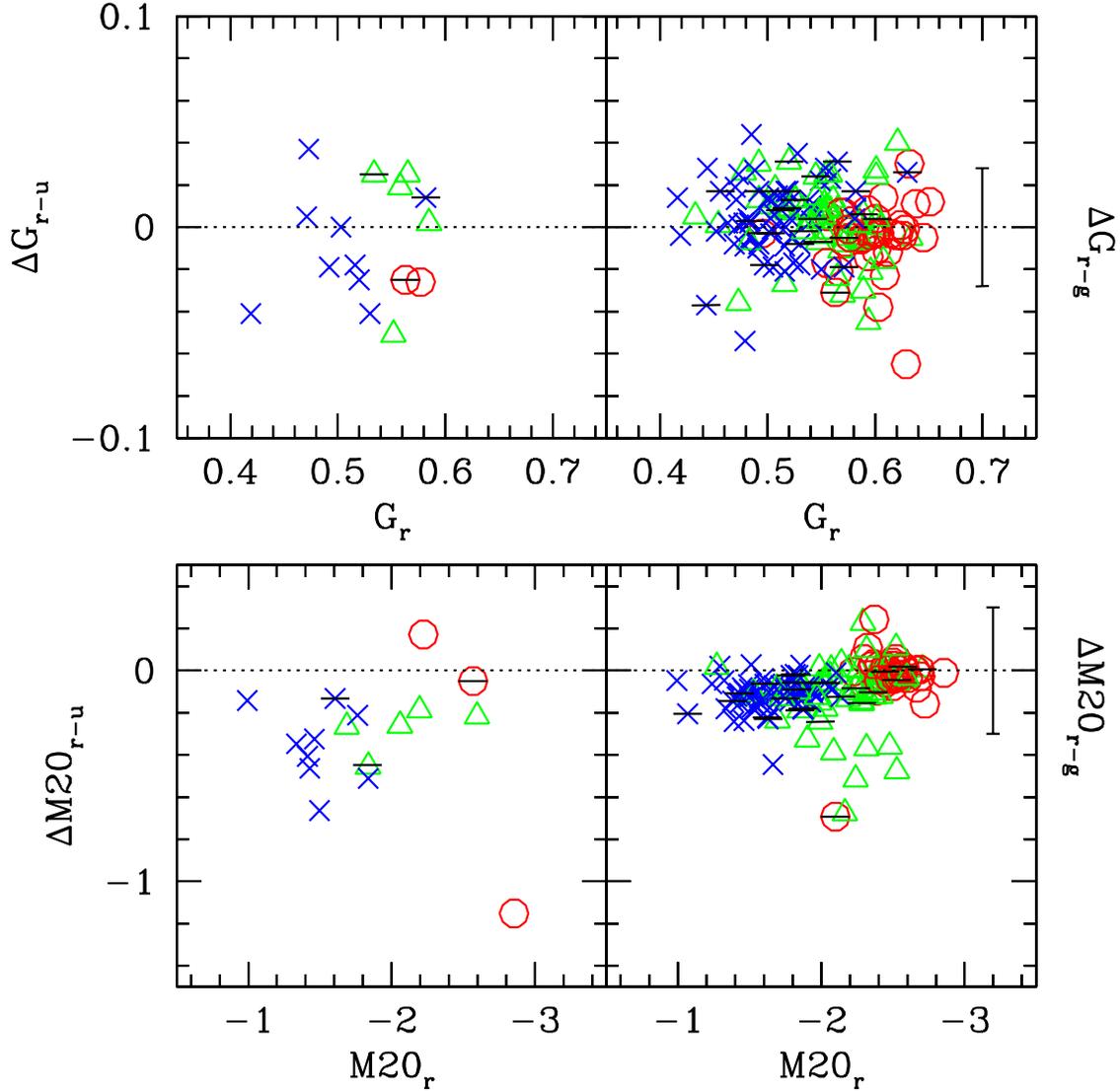}
\caption{ Change in $G$, $M_{20}$ morphology from $\sim$ 6500\AA\ ($R/r$) to $\sim$ 3600\AA\ ($u$) 
for SDSS u-selected sample with $S/N_u > 2.0$ (left)
and to $\sim$ 4500\AA\ ($B/g$) for Frei and SDSS galaxies with $S/N > 2.0$ (right). Error-bars and
point symbols are same as Figure 7.}
\end{figure}

\clearpage
\begin{figure}
\epsscale{1.0}
\plottwo{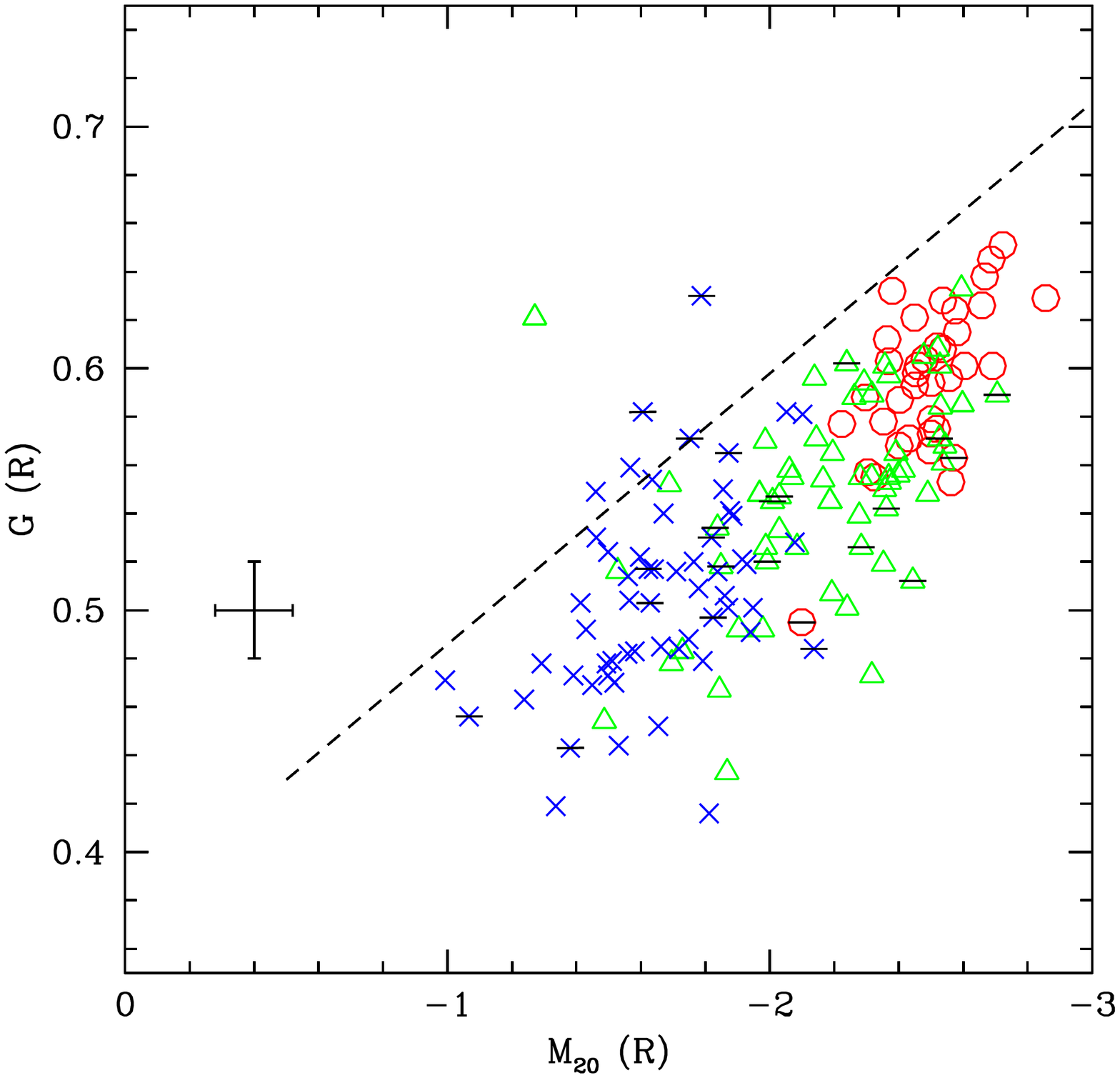}{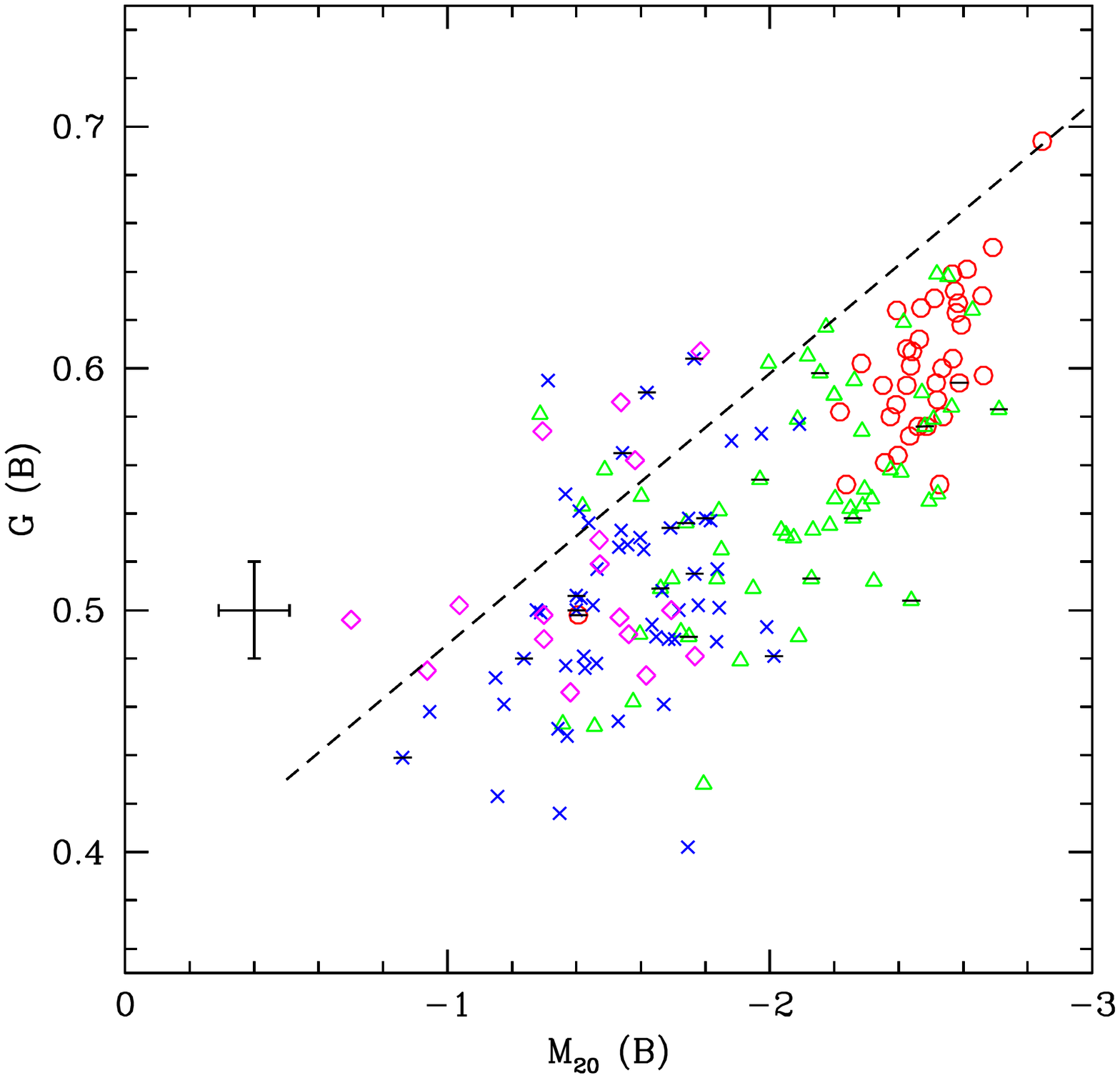}
\caption{$M_{20}$ v. $G$ for rest-frame $\sim 6500$\AA\ (left) and 4400\AA\ (right) observations of local galaxies
(red circles:E/S0, green triangles:Sa-Sbc, blue crosses:Sc-Sd, diamonds:dI, bars:edge-on spirals).
The error-bars are mean difference in $G$ and $M_{20}$ between SDSS $r$-band and Frei $R/r$ observations of
the same objects. Almost all the ``normal'' galaxies lie below the dashed line in the $R$-band plot. 
The outlying Sb galaxy NGC5850
has a strong star-forming ring and is in a close pair with NGC 5846.  Three of the outlying
dI in the $B$-band plot are star-bursting.}
\end{figure}

\clearpage
\begin{figure}
\plotone{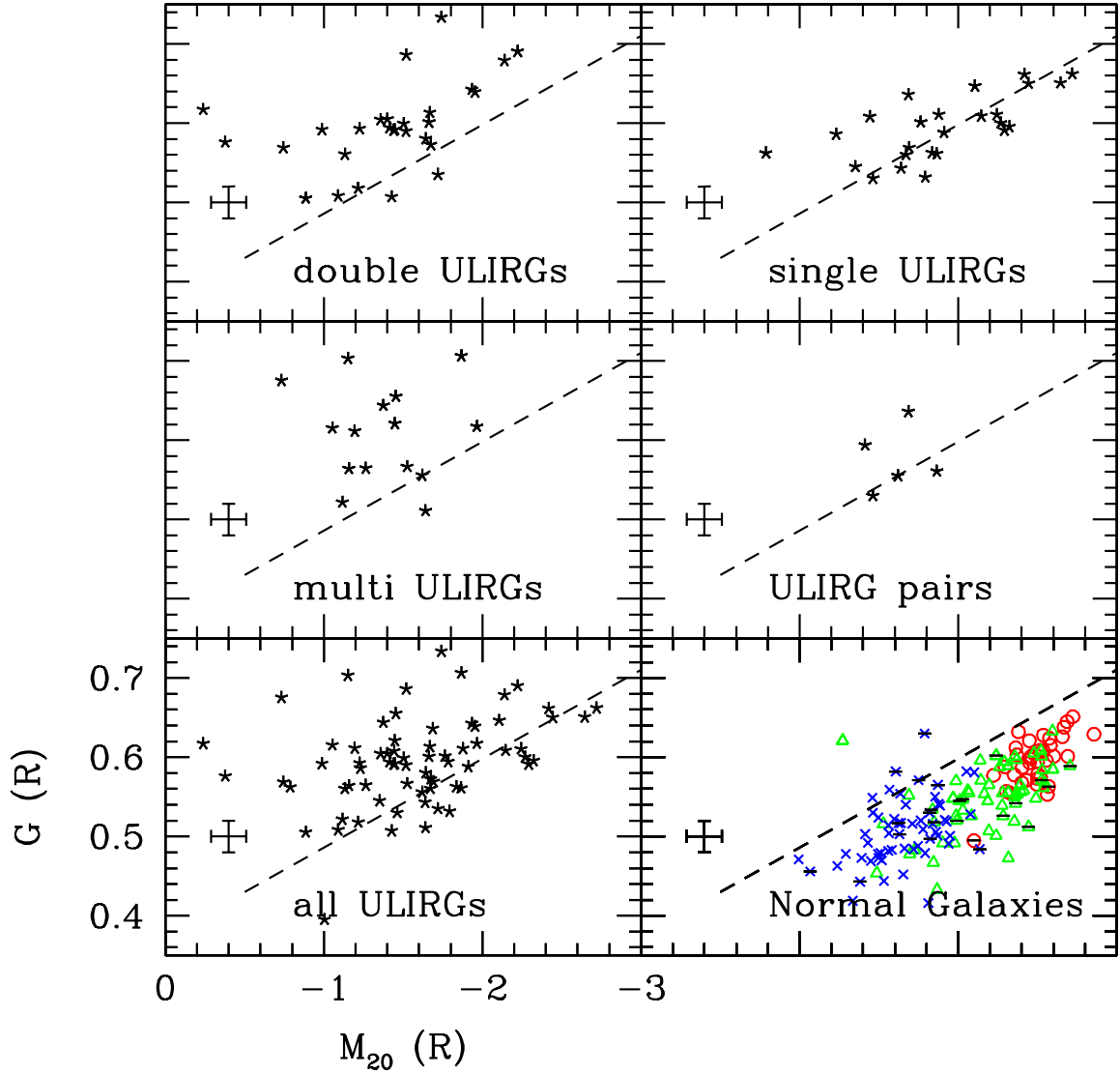}
\caption{$M_{20}$ v. $G$ for rest-frame $\sim 6500$\AA\ observations of local galaxies
(red circles:E/S0, green triangles:Sa-Sbc, blue crosses:Sc-Sd, stars:ULIRGs, bars:edge-on spirals).
The error-bars are mean difference in $G$ and $M_{20}$ between SDSS $r$-band and Frei $R/r$ observations of
the same objects. Almost all the ``normal'' galaxies lie below the dashed line.}
\end{figure}

\clearpage
\begin{figure}
\epsscale{1.5}
\plottwo{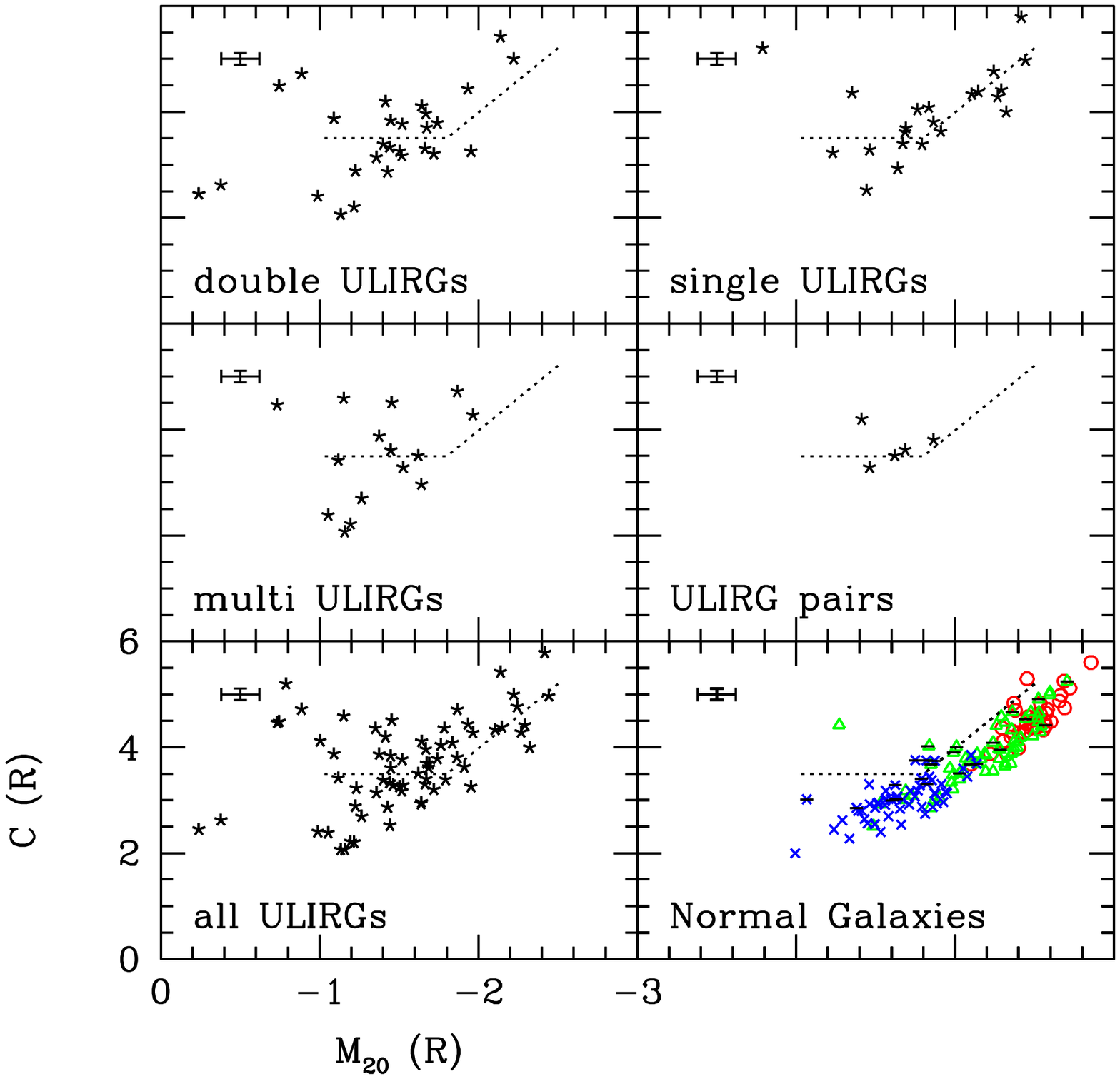}{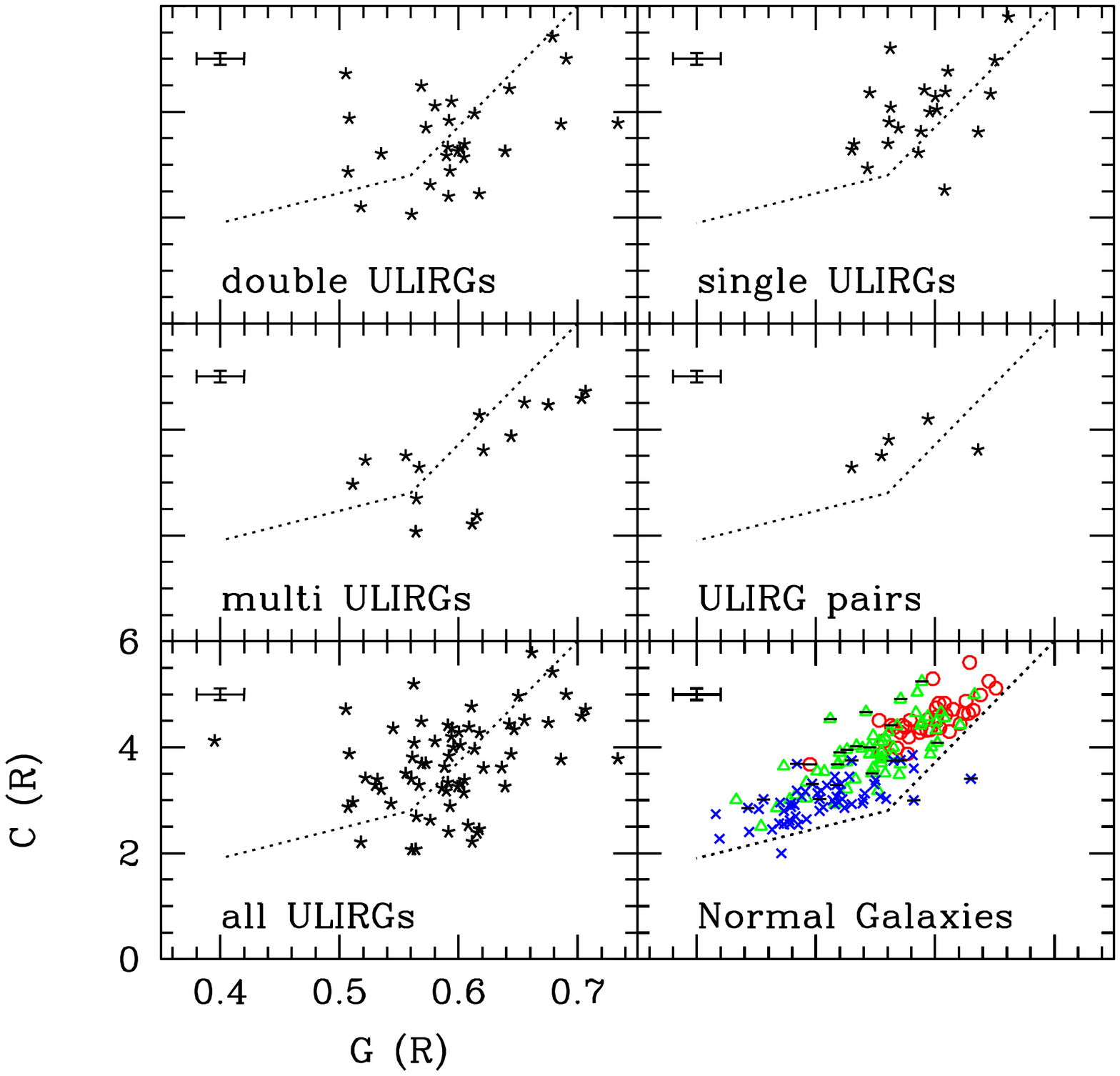}
\caption{$M_{20}$ v. $C$ and $G$  v. $C$ for rest-frame $\sim 6500$\AA\ observations of local galaxies.  Symbols
are same as Figure 10. The majority of normal galaxies lie to one side of the dashed lines.}
\end{figure}

\clearpage
\begin{figure}
\epsscale{1.5}
\plottwo{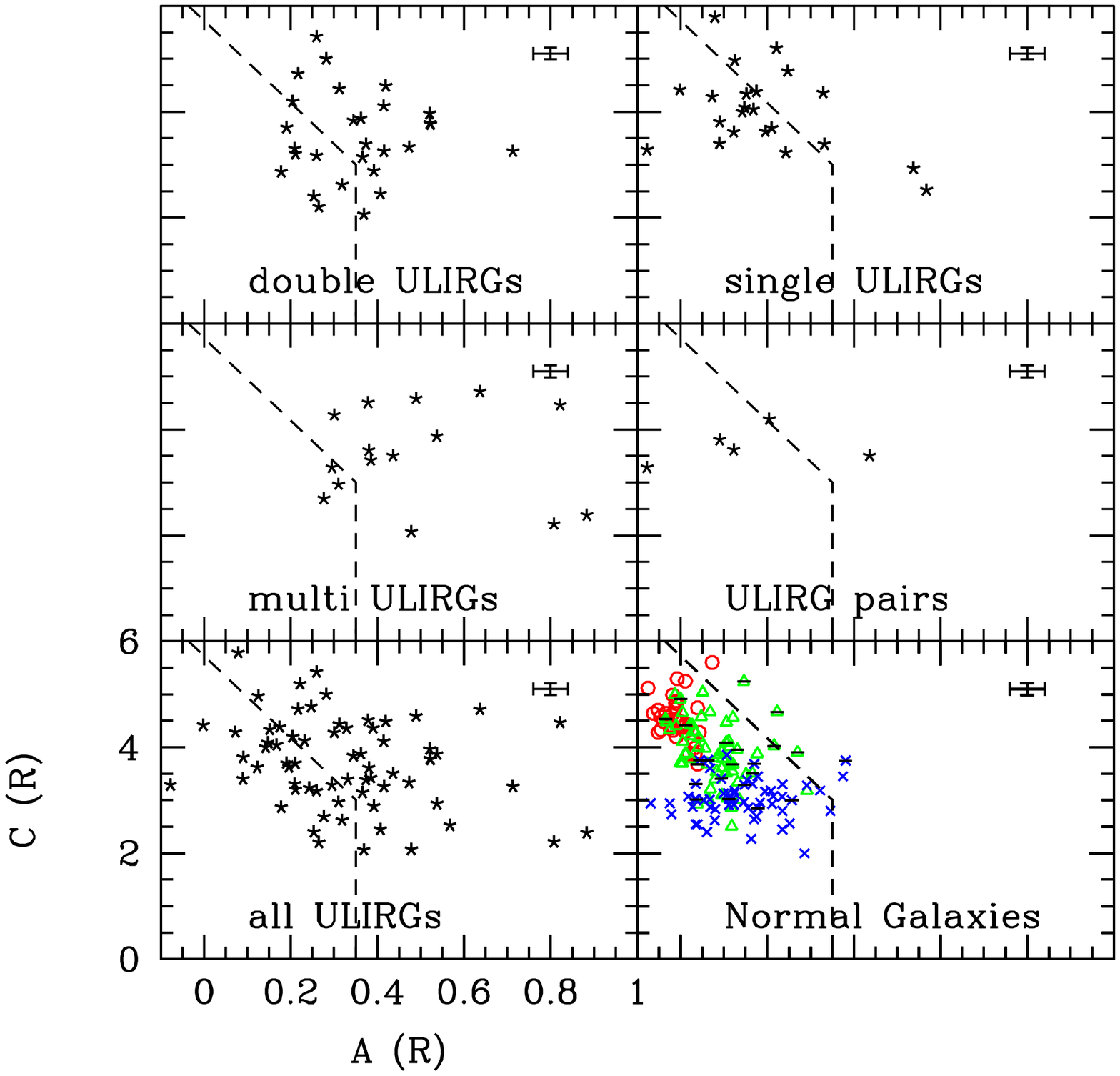}{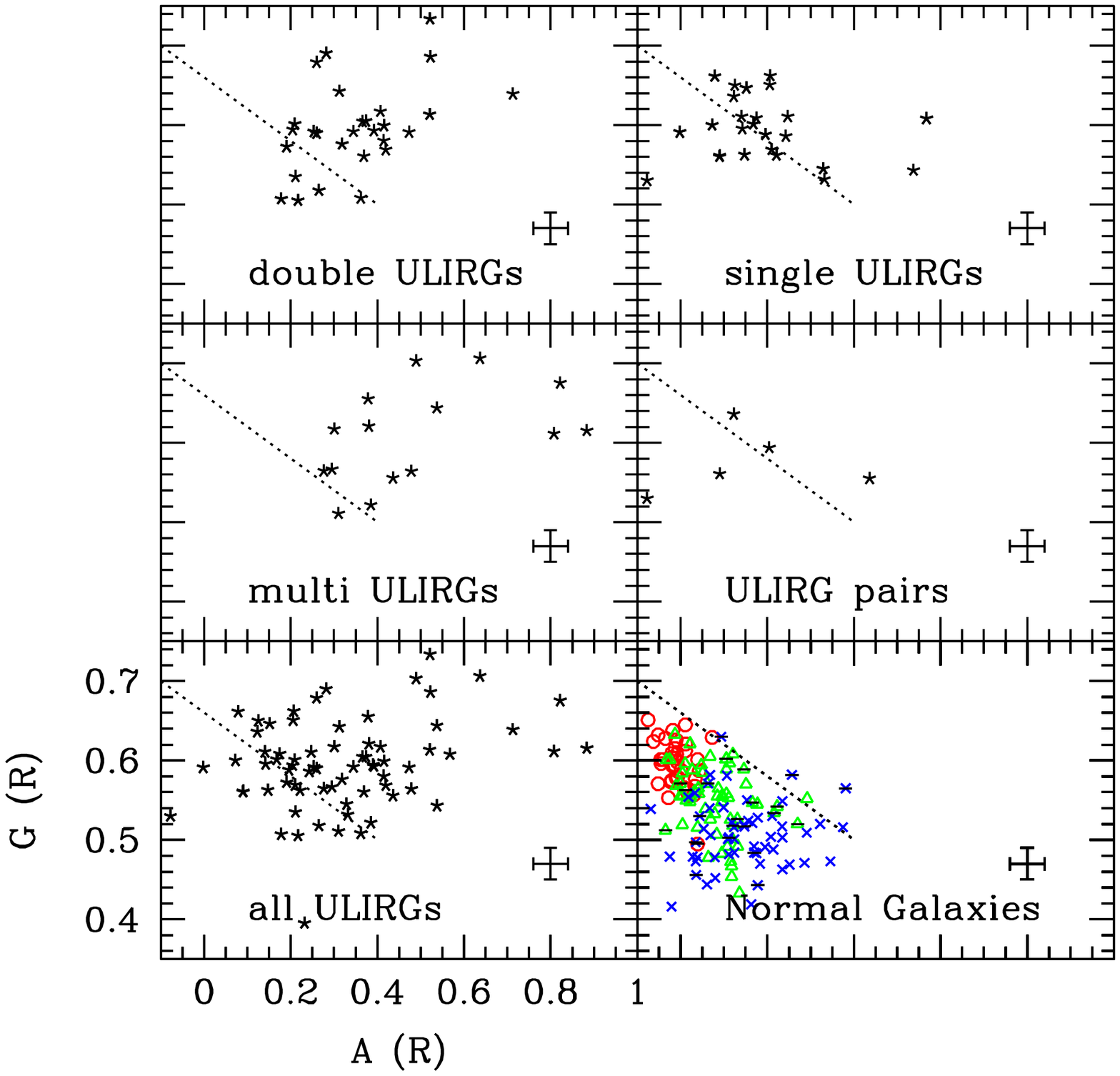}
\caption{$A$ v. $C$ and $G$ for rest-frame $\sim 6500$\AA\ observations of local galaxies. Symbols
are same as Figure 10. The majority of normal galaxies lie to one side of the dashed lines.}
\end{figure}

\clearpage
\begin{figure}
\epsscale{1.5}
\plottwo{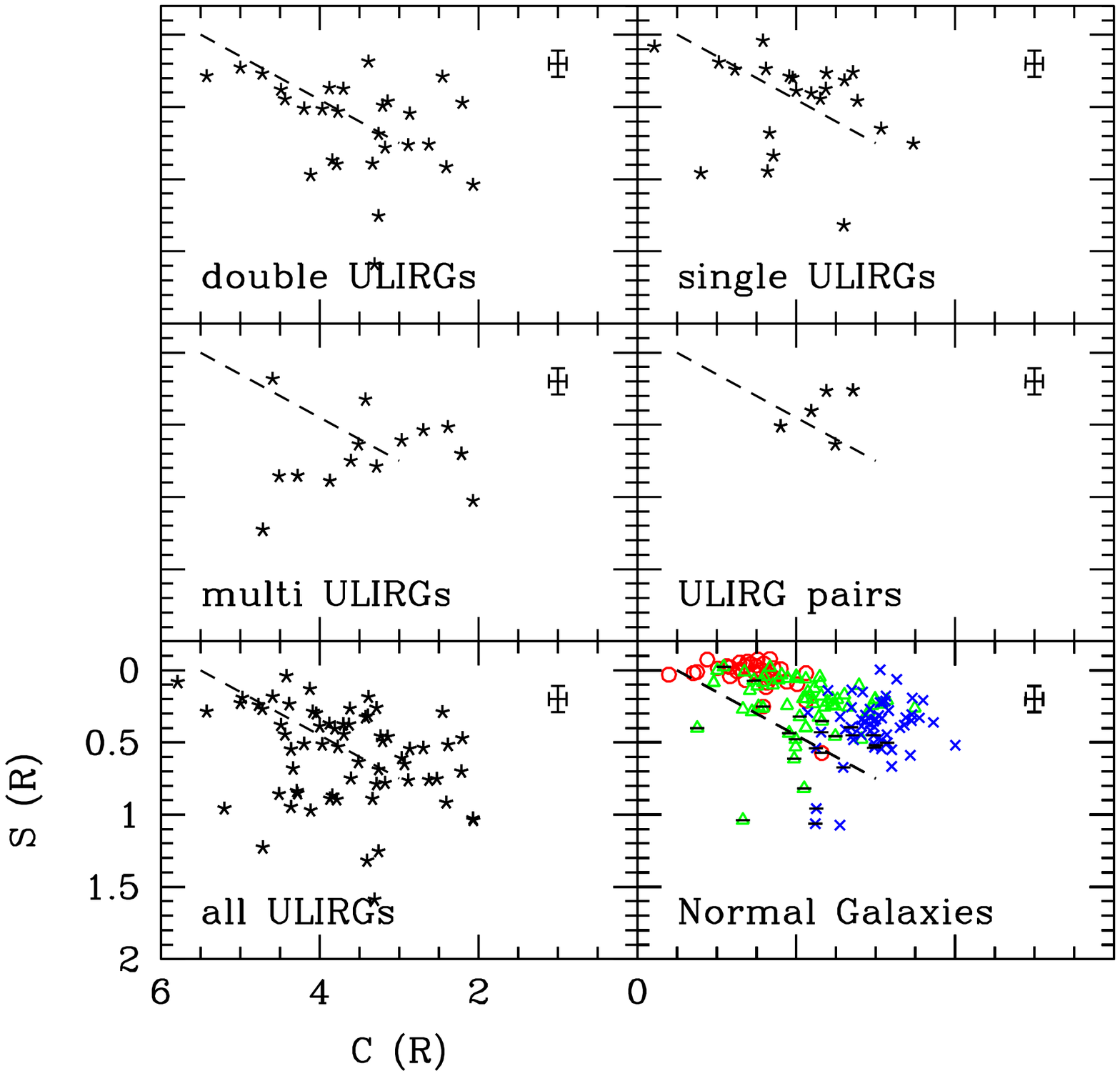}{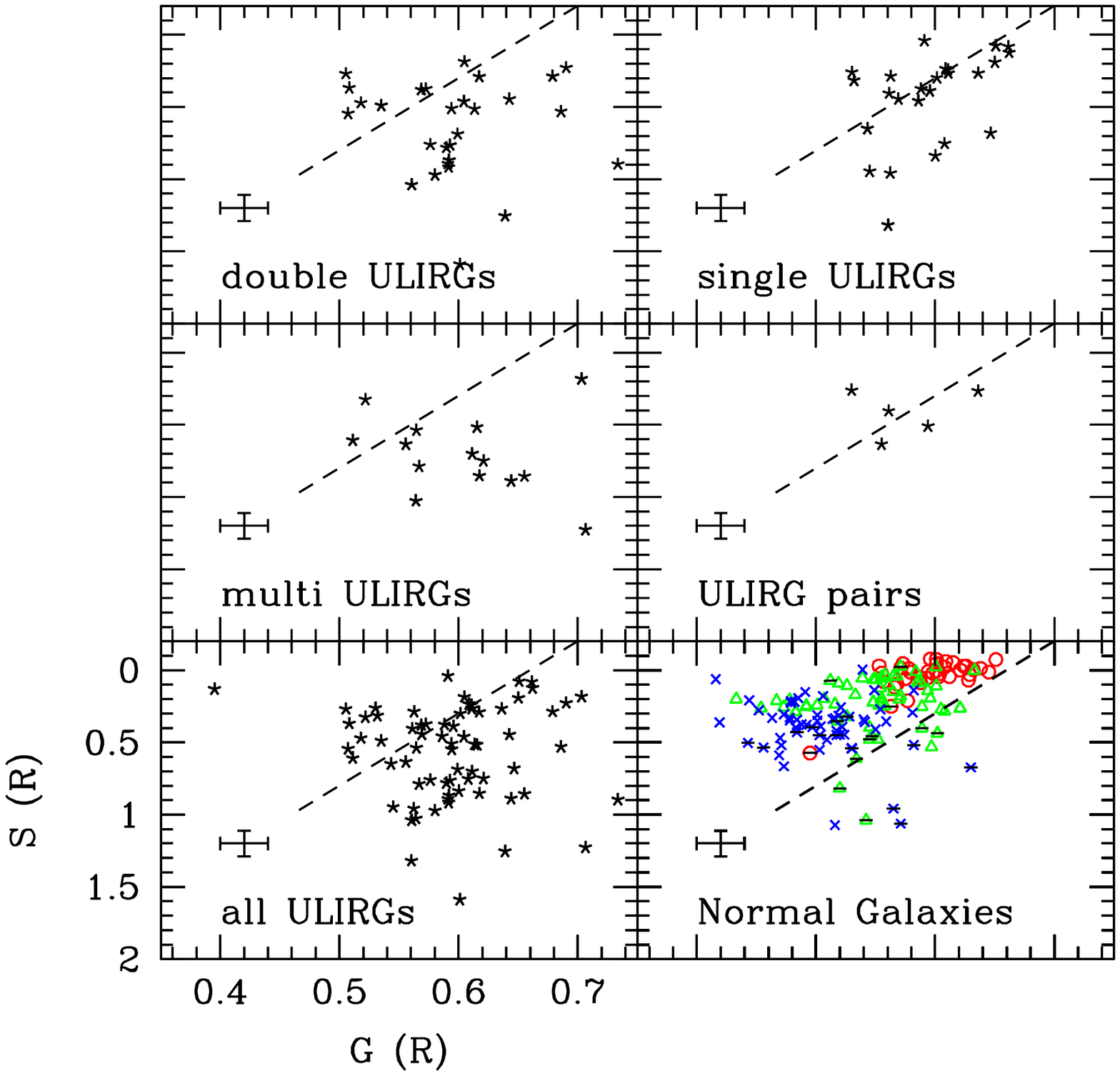}
\caption{$S$ v. $C$ and $G$ for rest-frame $\sim 6500$\AA\ observations of local galaxies.  Symbols
are same as Figure 10.}
\end{figure}

\clearpage
\begin{figure}
\epsscale{1.5}
\plottwo{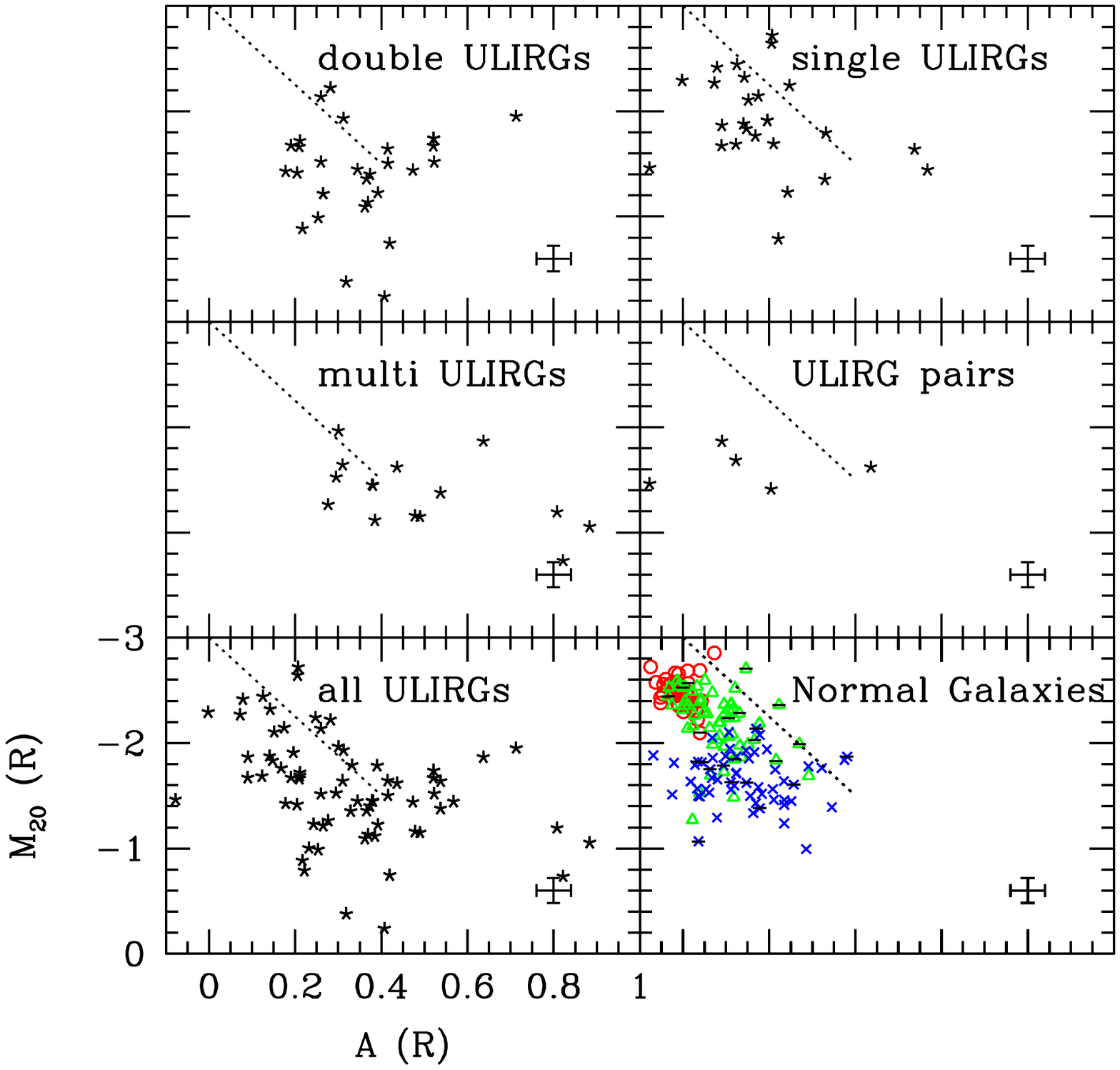}{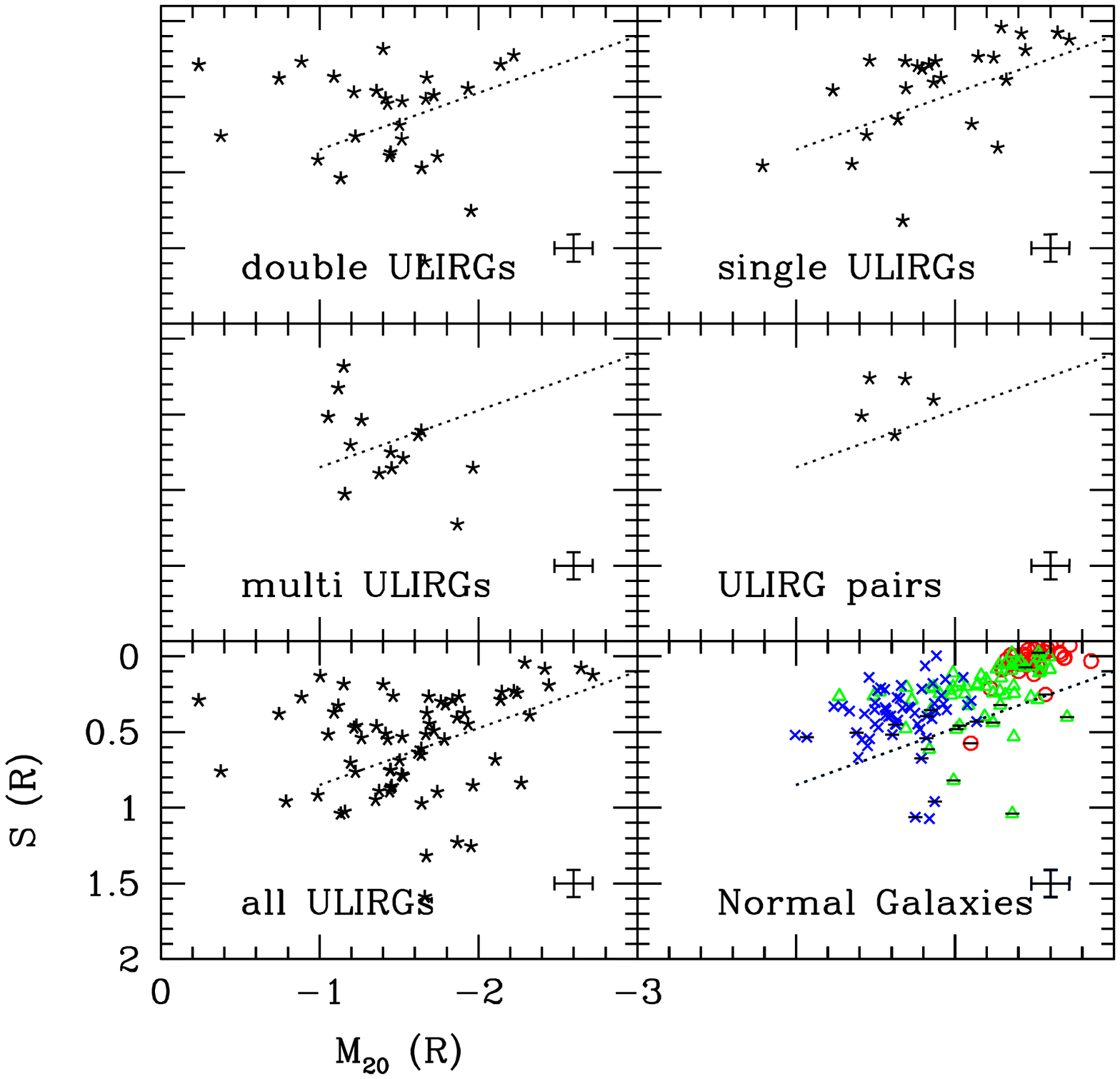}
\caption{$A$ v. $M_{20}$ and $M_{20}$ v. $S$ for rest-frame $\sim 6500$\AA\ observations 
of local galaxies.  Symbols are same as Figure 10.}
\end{figure}

\clearpage
\begin{figure}
\epsscale{1.0}
\plotone{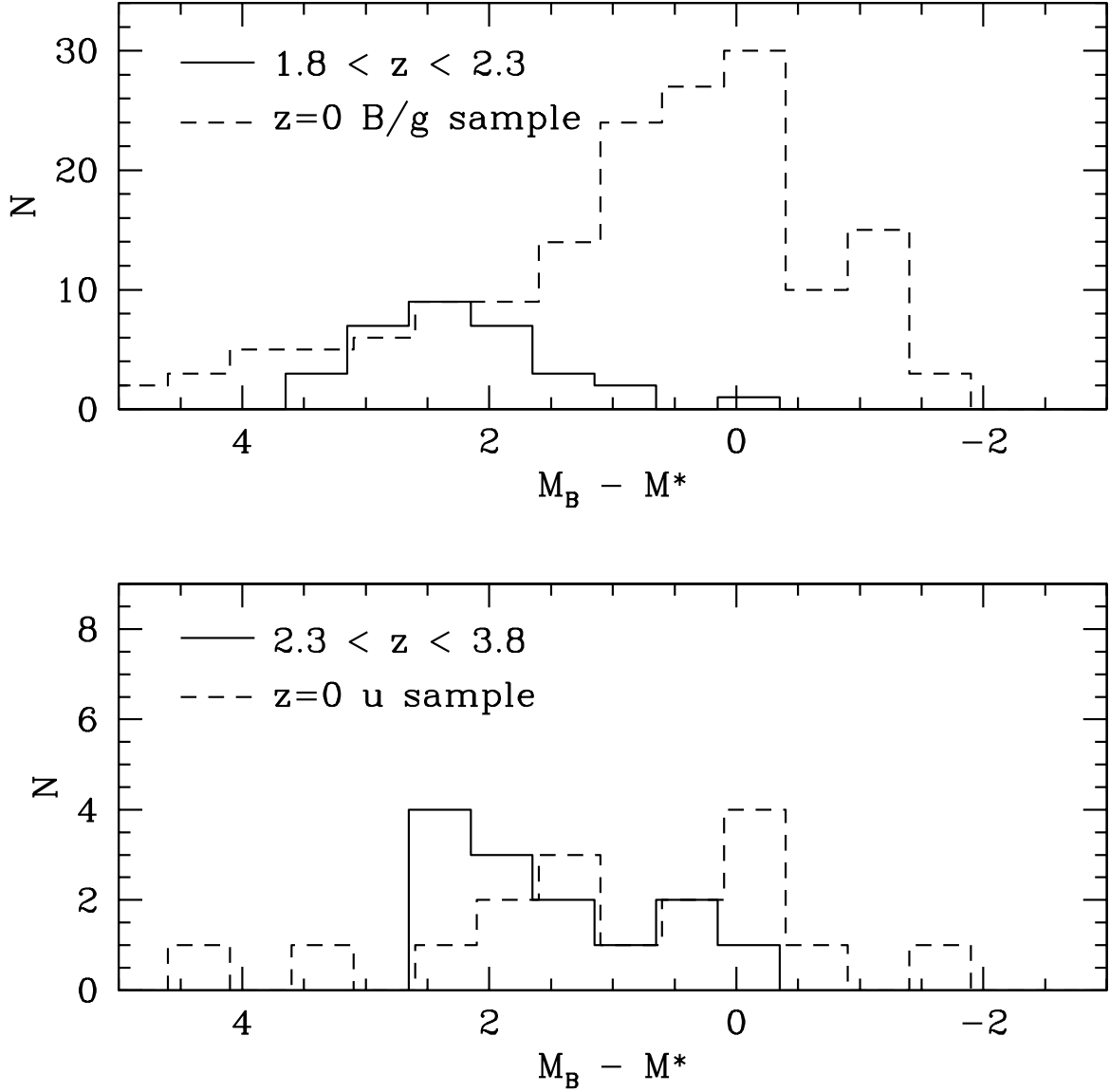}
\caption{ $M_B - M^*$ histograms for the $z \sim 2$ and $B/g$-band local galaxy samples (top)
and the $z \sim 3$ and $u$-band local galaxy samples (bottom).  $M^*$ is assumed to be $-20.1$
for the local galaxies and $-22.9$ for the $z \geq 2$ galaxies. \label{mbhist} }
\end{figure}

\clearpage
\begin{figure}
\epsscale{1.3}
\plottwo{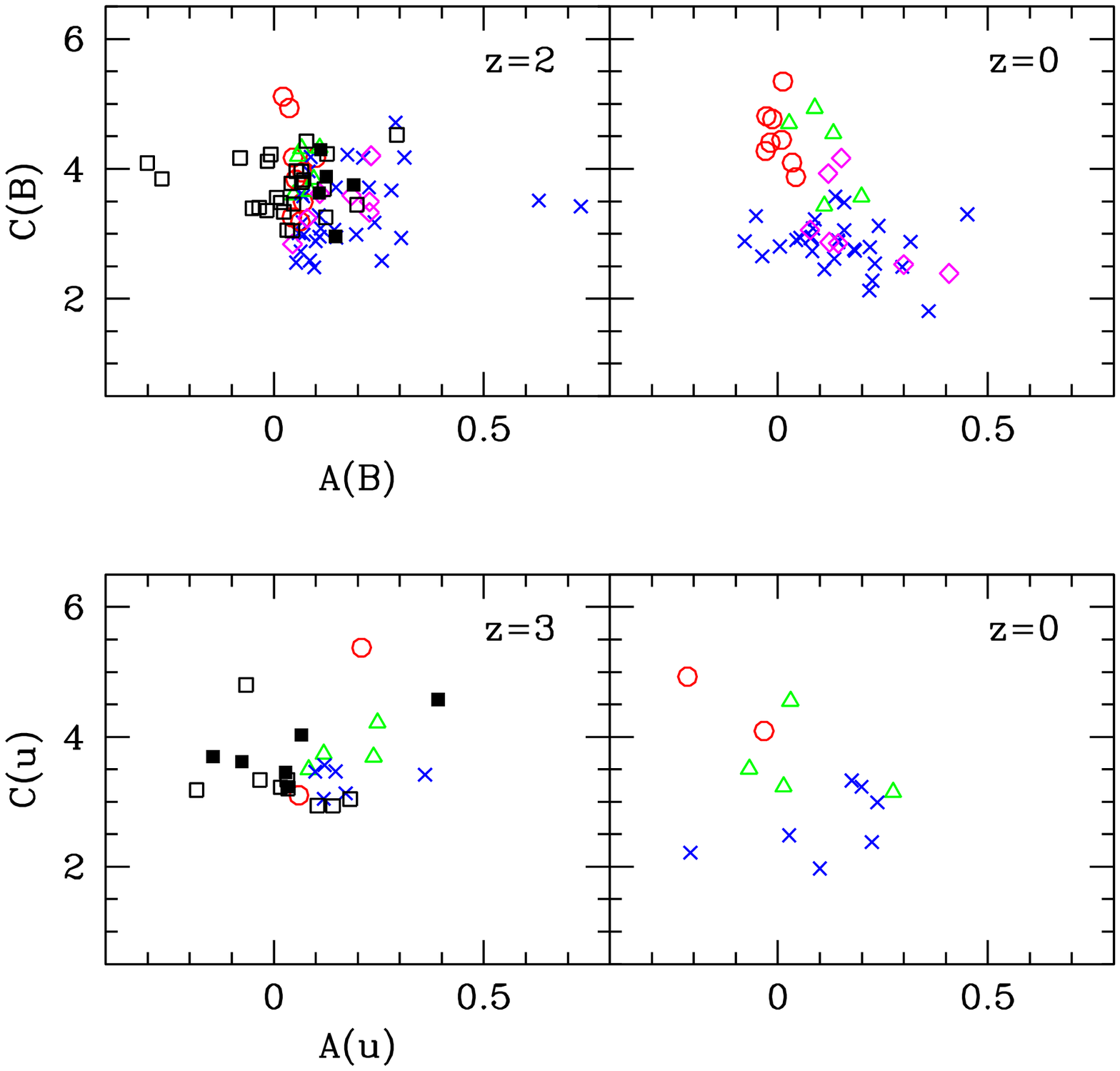}{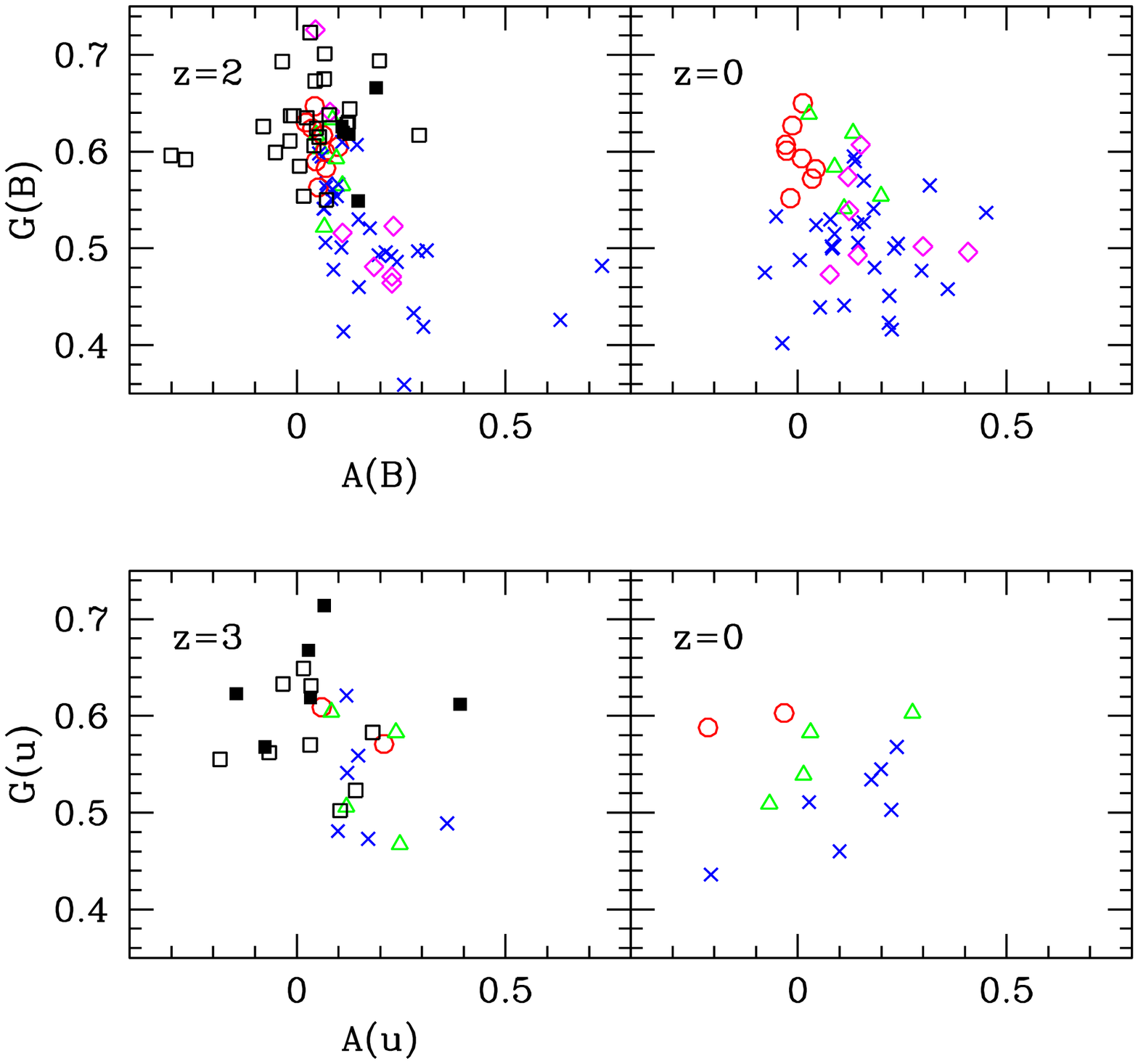}
\caption{ $A$ v. $C$ and $A$ v. $G$ for the HDFN Lyman break galaxies (open squares:
LBGs with spectroscopic redshifts, filled squares: LBGs with photometric redshifts).  The right
hand panels show observed morphologies of normal local galaxies (circles:E/S0, triangles:Sa-Sbc, crosses:Sc-Sd,
diamonds:dI).
The left hand panels show the observed LBG morphologies and the morphologies of local galaxies expected
for z=2-3 galaxies at the NICMOS HDFN image resolution. }
\end{figure}

\clearpage
\begin{figure}
\epsscale{1.3}
\plottwo{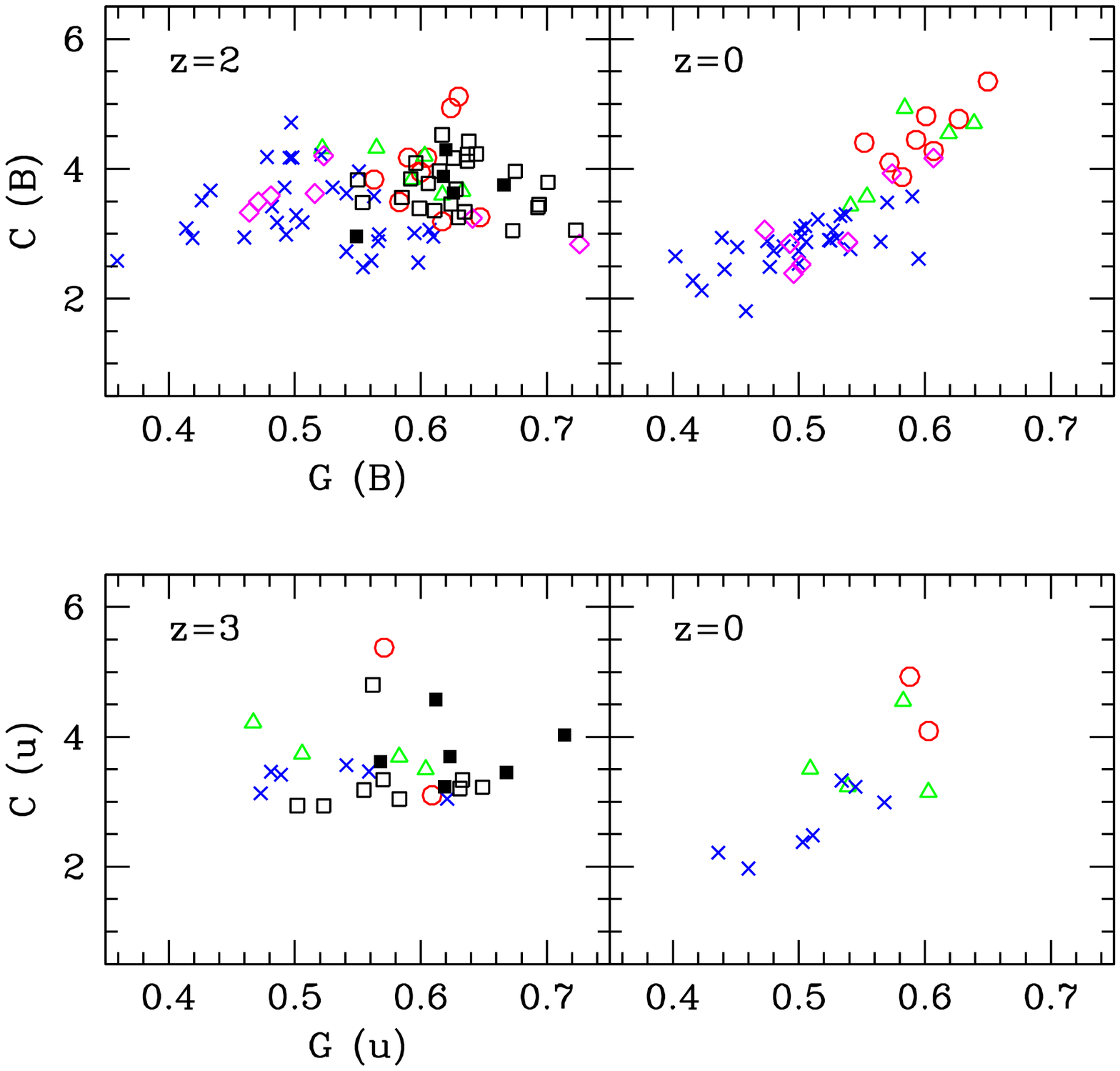}{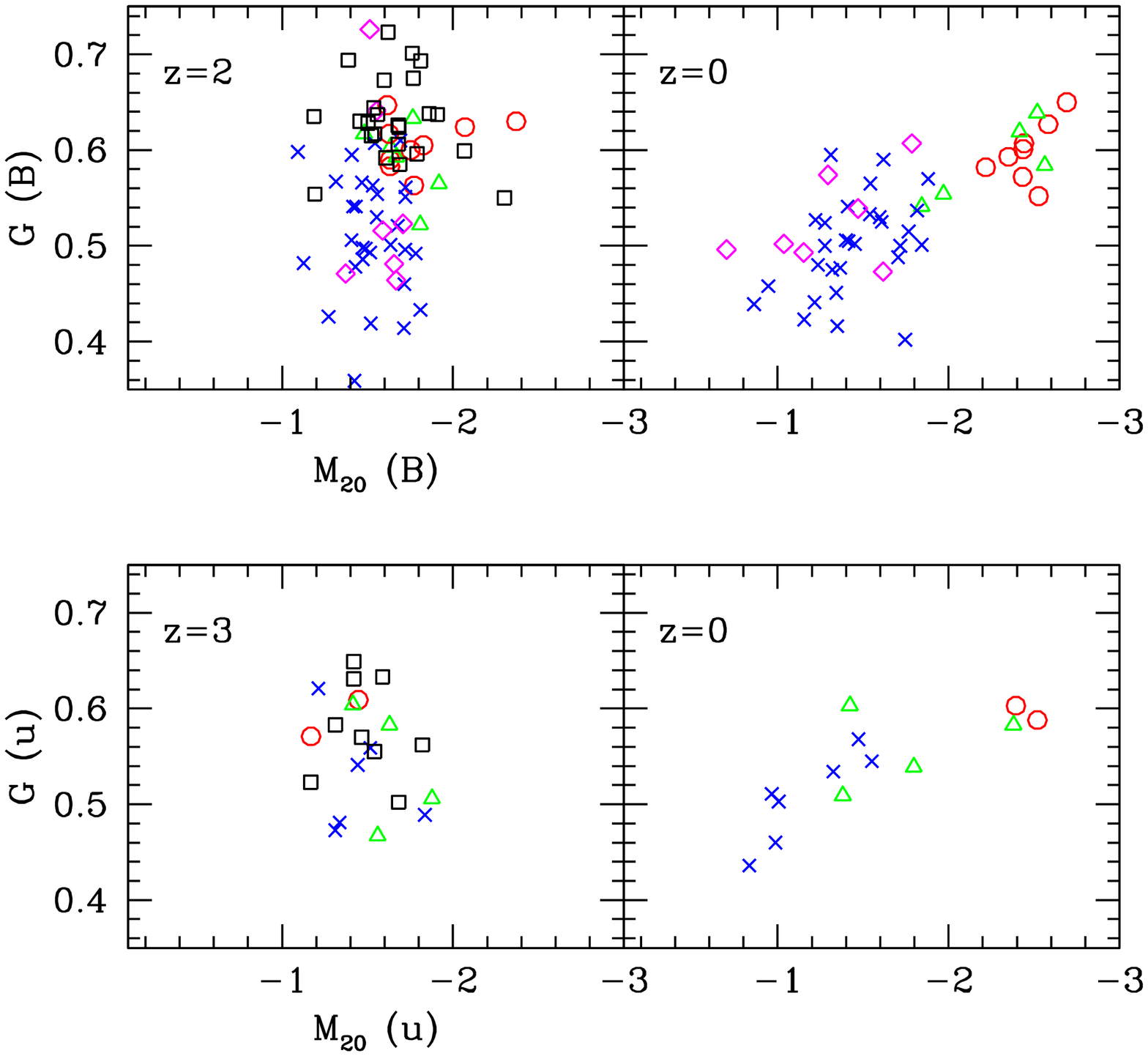}
\caption{$G$ v. $C$ and $M_{20}$ v. $G$ for the HDFN Lyman break galaxies (symbols are the same as Figure 16). }
\end{figure}

\end{document}